\documentclass[a4paper,11pt]{article}
\pdfoutput=1 

\usepackage{jheppub} 

\usepackage{gensymb}
\usepackage{subcaption}
\usepackage{soul}
\usepackage{slashed}
\usepackage{bm}
\usepackage{color}
\usepackage{tikz-feynman}
\usepackage{relsize}

\newcommand{\yN}{ \dfrac{Y_N}{Y_N^\text{eq}} }
\newcommand{\ySD}{ \dfrac{Y_{\Sigma \Delta} }{Y_{\Sigma \Delta}^\text{eq} } }
\newcommand{\ydD}{ \dfrac{Y_{\delta \Delta} }{Y_{\Sigma \Delta}^\text{eq} } }
\newcommand{\ydL}{ \dfrac{Y_{\delta L}}{Y_L^\text{eq}} }
\newcommand{\ydH}{ \dfrac{Y_{\delta H}}{Y_H^\text{eq}} }
\newcommand{\gDN}{\gamma_{D_N}}
\newcommand{\gDD}{\gamma_{D_\Delta}}
\newcommand{\rd}[4]{\gamma^{#1 #2}_{#3 #4} }

\newcommand{\rdDHbartoLbarN}{\gamma^{\Delta \overline{H}}_{\overline{L} N} }
\newcommand{\rdDNtoLbarH}{\gamma^{\Delta N}_{\overline{L} H} }
\newcommand{\rdDLtoHN}{\gamma^{\Delta L}_{H N} }
\allowdisplaybreaks[4]

\title{\boldmath Towards a more complete description of hybrid leptogenesis}

\author[a]{Rohan Pramanick,}
\author[a]{Tirtha Sankar Ray}
\author[b]{and Arunansu Sil}

\affiliation[a]{Department of Physics, Indian Institute of Technology Kharagpur, Kharagpur 721302, India}
\affiliation[b]{Department of Physics, Indian Institute of Technology Guwahati, Assam 781039, India}

\emailAdd{rohanpramanick25@gmail.com}
\emailAdd{tirthasankar.ray@gmail.com}
\emailAdd{asil@iitg.ac.in}

\abstract{Hybrid leptogenesis framework combining type I and type II seesaw mechanism for neutrino mass necessarily include scattering topologies involving both the scalar triplet and the right handed neutrino.  We demonstrate  that a systematic inclusion of these mixed scatterings can significantly alter the evolution of the number densities exhibiting upto a factor ten deviation in the predicted asymmetry as demonstrated by our benchmark scenarios.  We provide quantitative constraints on the degeneracy of the seesaw scales where the complete analysis becomes numerically significant, limiting the  validity of leptogenesis being dominated  by the lightest seesaw species only.}

\begin{document} 
\maketitle
\flushbottom

\section{Introduction} \label{sec:intro}

Generating the observed  Baryon Asymmetry of the Universe (BAU) through leptogenesis has been widely discussed in the literature  as it provides a  possible  connection between the eluding enigmas of the neutrino mass and the matter-antimatter asymmetry \cite{Fukugita:1986hr, Davidson:2008bu, DiBari:2021fhs}.  In this context the seesaw models that provide a natural framework for generating tiny masses for the neutrinos are of interest as they can also provide a handle to drive leptogenesis through CP violating out-of-equilibrium decays of the heavy seesaw states of different types \cite{King:2003jb, deGouvea:2016qpx, Cai:2017jrq, Ma:2006km}, viz. the right handed neutrino(s) for the Type I \cite{Mohapatra:1979ia, Schechter:1980gr}, scalar triplet(s) for Type II \cite{Lazarides:1980nt, Mohapatra:1986bd}  and the fermionic triplet(s) for Type III \cite{Foot:1988aq} seesaw mechanism.

In general if there are a number of heavy species with hierarchical masses that contribute to leptogenesis, the washout effects imply that the final asymmetry is dominated by the lightest species, see \cite{Engelhard:2006yg} for exceptions. Interestingly as the scales start coming close, processes that involve the participation of more than one type of these heavy species becomes important (henceforth called mixed processes) and consequently, estimation of the asymmetry requires a detailed study of the coupled Boltzmann equations including all the relevant states. In this work we present a hybrid scenario with Type I and Type II seesaw states simultaneously contributing to the neutrino mass and leptogenesis. We focus on the region of parameter space, in consonance with the neutrino oscillation data, where the mixed processes involving states belonging to the two different seesaw frameworks become numerically significant. We show that the full analysis tracking the number densities of all relevant species can significantly differ from the approximate results obtained by assuming that the dominant role is played by the lightest state.

The minimal version of the hybrid leptogenesis framework considered in this work, extends the Standard Model (SM) by one $SU(2)_L$ triplet scalar having hypercharge $Y = 1$, while the fermionic sector is augmented with a single right handed neutrino (RHN). The appearance of both RHN(s) and triplet(s) occur naturally in several extensions of SM such as the left-right symmetric model \cite{Pati:1974yy, Mohapatra:1974gc} and grand unified theories \cite{Georgi:1974sy, Georgi:1974yf, Fritzsch:1974nn, Langacker:1980js, Ibarra:2018dib, Bhattacharya:2021jli}.  The seesaw neutrino masses in agreement with the experimental limits can be generated by combining the type I and type II seesaw contributions \cite{Esteban:2020cvm}. The minimality of the framework imply that the CP violating decays of heavy state(s) necessarily require the simultaneous involvement of the scalar triplet as well the right handed neutrino. Leptogenesis in hybrid seesaw frameworks \cite{Mohapatra:1980yp, ODonnell:1993obr, Lazarides:1998iq} has been discussed in the literature mostly assuming hierarchical seesaw scales where leptogenesis is dominantly driven by decay of the lightest states \cite{Babu:2005bh, Akhmedov:2006yp, Rink:2020uvt, Hambye:2003ka, Datta:2021gyi}. However, in this work, we present for the first time a systematic analysis involving both the RHN and the scalar triplet contributing toward leptogenesis due to their closeness in mass-scales that finally leads to a substantial correction in final baryon asymmetry estimation. In particular, we show that there are new and modified $\Delta L = 1$ and $\Delta L = 2$ scattering processes in this simplified scenario of hybrid leptogenesis which are operative in the region where the type I and type II mass scales are relatively close and can significantly alter the baryon asymmetry parameter. As proof of principle, we present a couple of benchmark points to demonstrate this feature. We also provide a quantitative constraint on the closeness of mass scales in terms of a degeneracy parameter where these processes with mixed topologies become numerically significant.

The paper is organized as follows. In section \ref{sec:model} we introduce the minimal model where the neutrino masses are generated from a combination of type I and type II seesaw mechanisms. Leptogenesis within this hybrid framework is discussed in section \ref{sec: lepto}. We compare the complete analysis of the hybrid framework including the mixed processes with approximated solutions in section \ref{sec:num_analysis_results} and \ref{sec:validity_of_approximation} before concluding in section \ref{sec: conclusion}.

\section{The hybrid seesaw model and Neutrino mass} \label{sec:model}

In this section we present the minimal extension of the SM which generate neutrino masses and drive leptogenesis within the hybrid type I + II seesaw framework. The SM is extended with a $SU(2)_L$ singlet RHN $(N_R)$ and a complex scalar triplet $(\Delta)$ with hypercharge $Y = 1$ which are responsible to generate tiny neutrino masses via the type I and type II seesaw mechanisms respectively. The relevant part of the Yukawa Lagrangian includes
\begin{eqnarray} \label{eq:yukawa}
-\mathcal{L}_Y = {\mathcal{Y}_N}_{\alpha } \overline{L_\alpha} \widetilde{H} N_R + \dfrac{1}{2} {\mathcal{Y}_\Delta}_{\alpha \beta} \overline{L_\alpha^c} \Delta L_\beta + \dfrac{1}{2} M_N \overline{N_R}{N_R}^c + \text{ h.c.} \ ,
\end{eqnarray} 
where symbols have usual meaning and $\alpha, \beta = 1, 2, 3$ are flavour indices. The complex Yukawa couplings of the RHN denoted by ${\mathcal{Y}_N}_\alpha$ has $6$ independent real parameters, on the other hand for the scalar triplet the Yukawas are given by a $3 \times 3$ dimensional complex symmetric matrix $\mathcal{Y}_\Delta$ with $9$ independent elements. The scalar potential of this hybrid scenario is identical to a pure type II seesaw and {is represented} by
\begin{eqnarray} \label{eq:scalar_pot}
V (H, \Delta) &=& - \ m^2_H (H^\dag H)+\frac{\lambda }{4}(H^\dag H)^2 + M_\Delta^2 \text{Tr}(\Delta^\dag \Delta) + \lambda_2 \left[ \text{Tr}(\Delta^\dag \Delta)\right]^2 + \lambda_3 \text{Tr}(\Delta^\dag \Delta)^2  \nonumber \\
&& + \ \lambda_1 (H^\dag H) \text{Tr} (\Delta^\dag \Delta) + \lambda_4 (H^\dag \Delta \Delta^\dag H) + \mu \left[ H^\mathsf{T} i \tau_2 \Delta^\dag H + \text{ h.c.} \right] \ ,
\end{eqnarray}
where $\mu$ is the only dimensionful coupling present in the theory. After the electroweak symmetry breaking (EWSB), the Higgs acquires a vacuum expectation value $(v_H)$ which induces a tiny vev of the triplet given by $v_\Delta \sim \mu v_H^2/M_\Delta^2$ the value of which has an upper bound from precision electroweak measurements \cite{ParticleDataGroup:2022pth}.

Note that the bare mass term $(M_N)$ of $N_R$ present in eq. \ref{eq:yukawa} and the trilinear coupling $(\mu)$ of $\Delta$ given in eq. \ref{eq:scalar_pot} are chosen to be real without any loss of generality. The mass of the SM neutrinos are generated through a combination of type II and type I seesaw mechanisms given by
\begin{eqnarray} \label{eq:light_nu_mass_matrix}
M_\nu = M_\nu^{II} + M_\nu^{I} = \dfrac{1}{\sqrt{2}} \mathcal{Y}_\Delta v_\Delta - \dfrac{v_H^2}{2 M_N} \mathcal{Y}_N \mathcal{Y}_N^\mathsf{T} \ .
\end{eqnarray}

The neutrino mass matrix generated by a single RHN is inadequate in explaining both the solar and atmospheric mass differences. In this context, the inclusion of the triplet is a welcome feature. Contrarily, the type-II seesaw mechanism with a single scalar triplet although can satisfy the neutrino oscillation data by itself \cite{Esteban:2020cvm} (due to the presence of a larger number of free parameters in the Yukawa matrix $\mathcal{Y}_\Delta$), it fails to generate any CP asymmetry in the decay of the single $\Delta$. Thus, in the simplified setup, presence of both the heavy states is crucial to fulfill the necessary requirements to simultaneously satisfy the neutrino oscillation data and generation of lepton asymmetry, which we demonstrate in the following section \ref{sec: lepto}.

\section{Leptogenesis in the minimal hybrid seesaw model} \label{sec: lepto}

The study of Cosmic Microwave Background Radiation quantifies the present day abundance of baryon asymmetry as 
\begin{eqnarray}\label{bau}
&&Y_{\delta B} = 8.750\pm 0.077 \times 10^{-11} \ ,\label{YB}
\end{eqnarray}
observed by Planck \cite{Planck:2018vyg}. Such an asymmetry can be generated by various mechanisms such as GUT baryogenesis \cite{Ignatiev:1978uf,Yoshimura:1978ex,Toussaint:1978br,Dimopoulos:1978kv,Ellis:1978xg,Weinberg:1979bt,Yoshimura:1979gy,Barr:1979ye,Nanopoulos:1979gx,Yildiz:1979gx}, Affleck-Dine mechanism \cite{Affleck:1984fy,Dine:1995kz}, electroweak baryogenesis \cite{Rubakov:1996vz,Riotto:1999yt,Cline:2006ts} and baryogenesis via leptogenesis \cite{Riotto:1999yt,Pilaftsis:1997jf,Buchmuller:2004nz,Buchmuller:2005eh,Abada:2006ea,Davidson:2008bu,DiBari:2015oca}. We focus here on the leptogenesis scenario where a lepton asymmetry is created by the out of equilibrium decays of both the RHN and a scalar triplet in the context of a hybrid Type I + II scenario as sketched above, which finally would be converted into baryon asymmetry via the sphaleron process \cite{Khlebnikov:1988sr}. A systematic study of leptogenesis within this framework is discussed in the following subsections.

\subsection{CP asymmetry generation}

In the minimal hybrid seesaw construction, the involvement of two heavy seesaw states, $i.e.$ the RHN and the scalar, indicates that the initial lepton asymmetry can be contributed by the out of equilibrium decays of each of them in general.

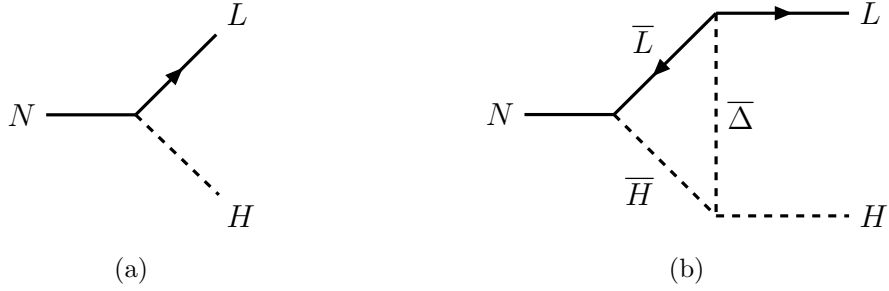
\begin{figure}[h!]
	\centering
	\begin{subfigure}{0.475\textwidth}
	\centering
	\scalebox{1}{
		\begin{tikzpicture}[baseline=(current bounding box.center)]
		\begin{feynman}
			\vertex (a) {\large $N$};
			\vertex [right = of a] (b);
			\vertex [above right=1.5cm of b] (c) {\large $L$};
			\vertex [below right=1.5cm of b] (d) {\large $H$};
			\diagram{
				(a) --[plain, very thick] (b) -- [fermion, very thick] (c);
				(b) --[scalar, very thick] (d);
					};
			
		\end{feynman}
		\end{tikzpicture}
	}
	\caption{}
	\label{fig:decay_tree_N}
	\end{subfigure}
	\begin{subfigure}{0.475\textwidth}
	\centering
	\scalebox{1}{
		\begin{tikzpicture}[baseline=(current bounding box.center)]
		\begin{feynman}
			\vertex (a) {\large $N$};
			\vertex [right = of a] (b);
			\vertex [above right = 1.9cm of b] (c);
			\vertex [below right = 1.9cm of b] (d);
			\vertex [right = 1.75cm of c] (e) {\large $L$};
			\vertex [right = 1.75cm of d] (f) {\large $H$};
			\diagram{
				(a) -- [plain, very thick] (b) -- [anti fermion, very thick, edge label = {\large  $\overline{L}$}] (c) -- [fermion, very thick] (e);
				(b) -- [scalar, very thick, edge label' = {\large $\overline{H}$}] (d) -- [scalar, very thick] (f);
(c) -- [scalar, very thick, edge label = {\large $\overline{\Delta}$} ] (d) ;
					};
			
		\end{feynman}
		\end{tikzpicture}
	}
	\caption{}
	\label{fig:decay_loop_N}
	\end{subfigure}
	\caption{Tree level (left) and triplet mediated one loop level (right) decay of RHN generating the asymmetry parameter $\epsilon_N$ .}
	\label{fig:decay_N}
\end{figure}

The RHN decays into a lepton and Higgs as denoted in figure \ref{fig:decay_N}. Crucially the interference between the tree level (figure \ref{fig:decay_tree_N}) and triplet mediated one loop level (figure \ref{fig:decay_loop_N}) diagrams generate the asymmetry given by \cite{Pramanick:2022put, Hambye:2003ka}
\begin{eqnarray} \label{eq:eps_N}
\epsilon_N &=& \sum_\alpha \dfrac{\Gamma(N \rightarrow L_\alpha + H) - \Gamma(N \rightarrow \overline{L_\alpha} + \overline{H})}{\Gamma(N \rightarrow L_\alpha + H) + \Gamma(N \rightarrow \overline{L_\alpha} + \overline{H})} \nonumber \\
&=& - \dfrac{1}{16 \pi^2 \Gamma_N} \sum_{\alpha, \beta} \mathfrak{Im} \big[ {\mathcal{Y}_N}_{\alpha} \ {\mathcal{Y}_N}_{\beta} \ {\mathcal{Y}_\Delta^*}_{\alpha \beta} \ \mu \big] \bigg[ 1 - \dfrac{M_\Delta^2}{M_N^2} \log \left( 1 + \dfrac{M_N^2}{M_\Delta^2} \right) \bigg] \ .
\end{eqnarray}

\begin{figure}[h!]
	\centering
	\begin{subfigure}{0.475\textwidth}
	\centering
	\scalebox{1}{
		\begin{tikzpicture}[baseline=(current bounding box.center)]
		\begin{feynman}
			\vertex (a) {\large $\overline{\Delta}$} ;
			\vertex [right = of a] (b) ;
			\vertex [above right = 1.5cm of b] (f1) {\large $L$} ;
			\vertex [below right = 1.5cm of b] (f2) {\large $L$} ;
			
			\diagram{
				(a) -- [scalar, very thick] (b) -- [fermion, very thick] (f1) ;
				(b) -- [fermion, very thick]	(f2)	 ;	
			} ;
			
		\end{feynman}
		\end{tikzpicture}
	}
	\caption{}
	\label{fig:decay_tree_D}
	\end{subfigure}
	\begin{subfigure}{0.475\textwidth}
	\centering
	\scalebox{1}{
	\begin{tikzpicture}[baseline=(current bounding box.center)]
		\begin{feynman}
			\vertex (a) {\large $\overline{\Delta}$};
			\vertex [right = of a] (b);
			\vertex [above right = 1.9cm of b] (c);
			\vertex [below right = 1.9cm of b] (d);
			\vertex [right = 1.75cm of c] (e) {\large $L$};
			\vertex [right = 1.75cm of d] (f) {\large $L$};
			\diagram{
				(a) -- [scalar, very thick] (b) -- [scalar, very thick, edge label = { $\overline{H}$}] (c) -- [fermion, very thick] (e);
				(b) -- [scalar, very thick, edge label' = {$\overline{H}$}] (d) -- [fermion, very thick] (f);
				(c) -- [plain, very thick, edge label = {\large $N$} ] (d) ;
					};
			
		\end{feynman}
		\end{tikzpicture}
	}
	\caption{}
	\label{fig:decay_loop_D}
	\end{subfigure}
	\caption{Tree level (left) and RHN mediated one loop level (right) decay of triplet generating the asymmetry parameter $\epsilon_\Delta$ .}
	\label{fig:decay_D}
\end{figure}
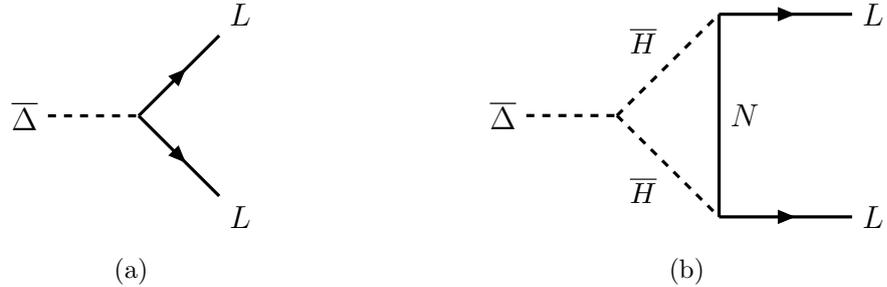

On the other hand, the CP asymmetry from the decay of the triplet results from the interference between the tree (figure \ref{fig:decay_tree_D}) and RHN mediated one loop (figure \ref{fig:decay_loop_D}) diagrams and is given by \cite{Pramanick:2022put, Hambye:2003ka}
\begin{eqnarray} \label{eq:eps_D}
\epsilon_\Delta &=& \sum_{\alpha, \beta} \dfrac{\Gamma(\overline{\Delta} \rightarrow L_\alpha + L_\beta) - \Gamma(\Delta \rightarrow \overline{L_\alpha} + \overline{L_\beta})}{\Gamma(\overline{\Delta} \rightarrow L_\alpha + L_\beta) + \Gamma(\Delta \rightarrow \overline{L_\alpha} + \overline{L_\beta})} \nonumber \\
&=&\dfrac{1}{64 \pi^2 \Gamma_\Delta} \sum_{\alpha, \beta} \mathfrak{Im} \big[ {\mathcal{Y}_N^*}_{\alpha} \ {\mathcal{Y}_N^*}_{\beta} \ {\mathcal{Y}_\Delta}_{\alpha \beta} \ \mu \big] \bigg[ \dfrac{M_N}{M_\Delta}  \log \left( 1 + \dfrac{M_\Delta^2}{M_N^2} \right) \bigg] \ .
\end{eqnarray}
In the above expression, $\Gamma_N$ and $\Gamma_\Delta$ denote the total decay width of RHN and triplet respectively and is given by
\begin{equation} \label{eq:decay_widths}
\Gamma_N = \dfrac{M_N}{8 \pi} \text{Tr} \left( {\mathcal{Y}_N}^\dagger \mathcal{Y}_N \right) \ ; \quad \Gamma_\Delta = \dfrac{M_\Delta}{8 \pi} \Bigg[ \text{Tr} \left( {\mathcal{Y}_\Delta}^\dagger \mathcal{Y}_\Delta \right) + \left( \dfrac{\mu}{M_\Delta} \right)^2 \Bigg] \ .
\end{equation}
\begin{figure}[t] 
	\centering
	\begin{subfigure}{0.49\textwidth}
		\centering
		\includegraphics[width=1\linewidth]{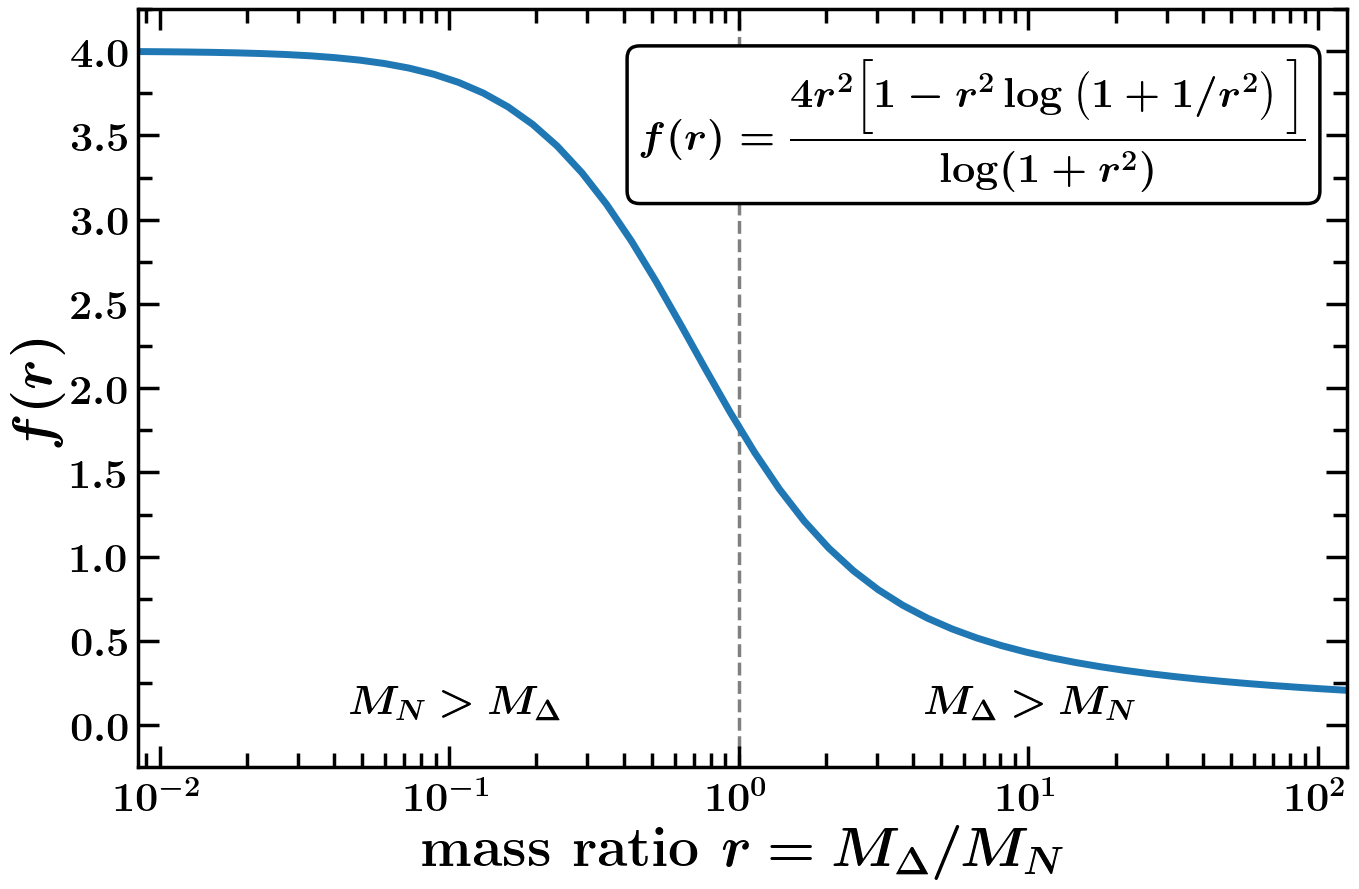} 
		\caption{}
		\label{fig:eps_r_part_f}
	\end{subfigure}
	\begin{subfigure}{0.49\textwidth}
		\centering
		\includegraphics[width=1\linewidth]{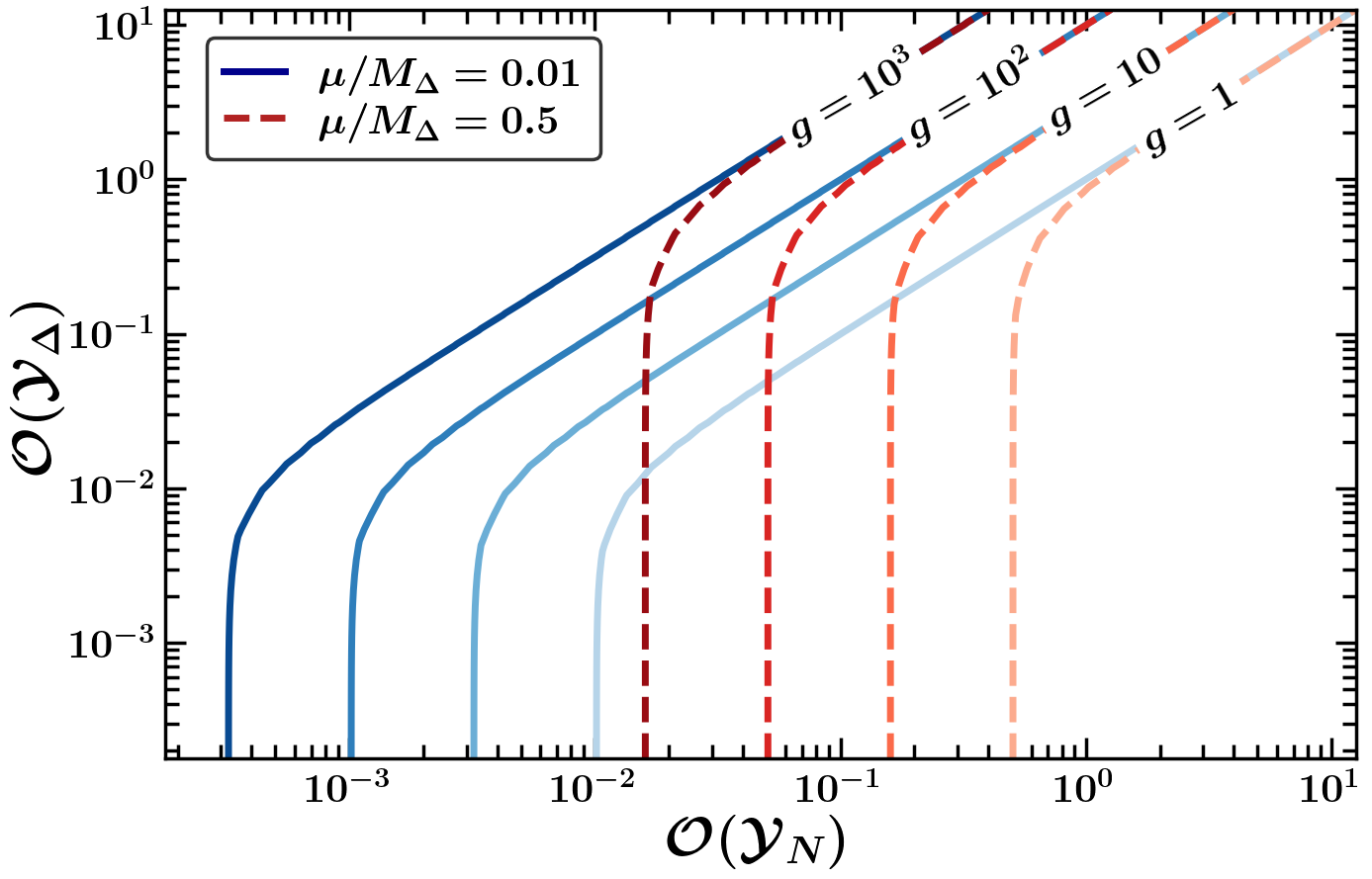} 
		\caption{}
		\label{fig:eps_r_part_g}
	\end{subfigure}
	\caption{Left panel shows the variation of $f(r)$ defined in eq. \ref{eq:eps_r} with respect to the mass ratio $r$ where mass degeneracy of RHN and triplet is denoted via the vertical gray dashed line. Right panel shows order of magnitude contours of $g(\mathcal{Y}_N, \mathcal{Y}_\Delta, \mu/M_\Delta)$ appearing in eq. \ref{eq:eps_r} in the plane of Yukawa couplings $\mathcal{O}(\mathcal{Y}_N)$ vs. $\mathcal{O}(\mathcal{Y}_\Delta)$ for fixed value of $\mu/M_\Delta = 0.01 (0.5)$ denoted by solid (dashed) blue (red) curves. }
\end{figure}

In this simplified setup, it is interesting to note that the combination of the neutrino Yukawas and trilinear couplings that appear in the expression of $\epsilon_N$ and $\epsilon_\Delta$ in eqs. \ref{eq:eps_N} and \ref{eq:eps_D} respectively are complex conjugates of each other. Thus the contribution of this specific combination cancels out in the expression of the ratio of the CP asymmetry parameters defined as $\epsilon_r = \epsilon_N/\epsilon_\Delta$, representative of a relative hierarchy between the CP asymmetries generated from RHN and Triplet decays. Apart from the couplings involved, the triplet to RHN mass ratio defined by $r = M_\Delta/M_N$ also plays a significant role in determining $\epsilon_r$. In order to extract the effects of various couplings and masses in the asymmetry parameter, the ratio $\epsilon_r$ can be factorized as 
\begin{equation} \label{eq:eps_r}
\begin{aligned}
\epsilon_r &= \Big[\dfrac{4 r^2 \left(1 - r^2 \log ( 1 + 1/r^2) \right)} {\log (1 + r^2)} \Big] \Big[ \dfrac{ \text{Tr} \left(\mathcal{Y}_\Delta^\dagger \mathcal{Y}_\Delta \right) + \left(\dfrac{\mu}{M_\Delta} \right)^2 }{ \text{Tr} \left( \mathcal{Y}_N^\dagger \mathcal{Y}_N \right)}\Big] \ , \\
&\equiv f(r) \ g(\mathcal{Y}_N, \mathcal{Y}_\Delta, \mu/M_\Delta) \ ,
\end{aligned}
\end{equation}
where the functions $f(r)$ and $g(\mathcal{Y}_N, \mathcal{Y}_\Delta, \mu/M_\Delta)$ encapsulate the dependence on mass ratio $r = M_\Delta/M_N$ and Yukawa couplings respectively. In order to exhibit such dependence more explicitly, we include figures \ref{fig:eps_r_part_f} and \ref{fig:eps_r_part_g}. As can be seen from it, the mass ratio dependent factor $f(r)$ appearing in eq. \ref{eq:eps_r} attains a maximum value $4.0$ corresponding to a smaller value of $r \sim \mathcal{O}(10^{-2})$ while it asymptotically approaches zero for larger value of $r \gg 1$. On the other hand, a contour plot for function $g$ is drawn in $\mathcal{Y}_N - \mathcal{Y}_\Delta$ plane for some fixed values of $g$ and $\mu/M_\Delta$ in figure \ref{fig:eps_r_part_g}. As evident, a relatively large (or small) value of $g$ or initial relative asymmetry between RHN or triplet decay is obtainable by appropriately tuning the quantities $\mu/M_\Delta$ and/or hierarchy between couplings $\mathcal{Y}_N$ and $\mathcal{Y}_\Delta$. From figures \ref{fig:eps_r_part_f} and \ref{fig:eps_r_part_g} it is easy to read off $f$ and $g$ for any generic hybrid scenario and estimate the initial asymmetries from eq. \ref{eq:eps_r}.

\subsection{Evolution of baryon asymmetry} 

In this section we discuss the evolution of how the primordial lepton asymmetry can be associated with the present day observation of BAU \cite{Planck:2018vyg}. In order to investigate the evolution of $B - L$ asymmetry through the out of equilibrium decay of heavier species, we need to solve the set of coupled Boltzmann equations (BEs) associated with abundances of various particles. This $B - L$ asymmetry can then be transferred to a net baryon asymmetry through non perturbative sphaleron processes \cite{Khlebnikov:1988sr}. The sphaleron factor in this model is calculated to be $a_\text{sph} = Y_{\delta B}/Y_{\delta (B - L)} = 12/37$ assuming sphaleron remains at equilibrium after EW phase transition \cite{Harvey:1990qw}.

We consider Yukawa couplings in the range of $\mathcal{O}(10^{-3} - 10^{-1})$ for both RHN and scalar triplet while being in alignment with the \emph{seesaw} spirit which fixes the mass of the heavy particles to be greater than $10^{12}$ GeV. This high scale of leptogenesis can be safely considered to be unflavoured\footnote{The reheating temperature is expected to be accordingly higher here so that the decay of the heavy seesaw states take place in radiation dominated era. For the  decay happening in an extended reheating era, the conventional flavor regimes of leptogenesis would be modifed \cite{Datta:2022jic}.} as none of the SM Yukawa interactions enter thermal equilibrium at these temperatures. 
In order to write down the BEs for this mixed seesaw framework, one needs to track the abundances of RHN, triplet (both its symmetric and asymmetric part denoted by $\Sigma \Delta = \Delta + \overline{\Delta}$ and $\delta \Delta = \Delta - \overline{\Delta}$ respectively) and the $B-L$ asymmetry which form a set of four coupled differential equations given by
\begin{subequations}\label{eq:boltz_full}
\begin{align}
\dot{Y}_N = {}& - \left(\yN - 1 \right) \Bigg[ \gDN + 2 \Big( \rd{N}{L}{\overline{Q}}{t_R}  + \rd{N}{Q}{L}{t_R} + \rd{N}{\overline{t_R}}{L}{\overline{Q}} \Big) \Bigg]  \nonumber \\
& - \ 2 \Bigg[ \left( \yN \ySD - 1 \right) \rdDNtoLbarH + \left( \yN - \ySD \right) \bigg(\rdDHbartoLbarN + \rdDLtoHN \bigg) \Bigg] \ , \\
\dot{Y}_{\Sigma \Delta} = {}& - \left( \ySD - 1 \right) \gDD - 2 \left\lbrace \left( \ySD \right)^2 - 1 \right\rbrace \gamma_A \nonumber \\
& - \ 2 \Bigg[ \left( \yN \ySD - 1 \right) \rdDNtoLbarH + \left( \ySD - \yN \right) \bigg( \rdDHbartoLbarN + \rdDLtoHN \bigg) \Bigg] \ , \\
\dot{Y}_{\delta \Delta} = {}& - \left( \ydD + B_L \ydL - B_H \ydH \right) \gDD \nonumber \\
& - \Bigg[ \left( 2 \ydD \yN + \ydL - \ydH \right) \rdDNtoLbarH + \left( 2\ydD - \ydH + \ydL \yN  \right) \rdDHbartoLbarN \nonumber \\
& \hspace{0.4cm} + \left( 2\ydD + \ydL - \ydH \yN \right) \rdDLtoHN \Bigg] \ , \\
\dot{Y}_{\delta (B - L)} = {}& - \left(\yN - 1 \right) \epsilon_N \gDN + 2 \left( \ydL + \ydH \right) \gDN \nonumber \\
& - \dfrac{1}{2} \left(\ySD - 1 \right) \epsilon_\Delta \gDD + B_L \left( \ydD + \ydL \right) \gDD \nonumber \\
& + \ydL \Bigg[ \left( \rd{N}{Q}{L}{t_R} + \rd{N}{\overline{t_R}}{L}{\overline{Q}} \right) - \yN \ \rd{N}{L}{\overline{Q}}{t_R} \Bigg] + 2 \left( \ydL + \ydH \right) \Big( {\gamma_\text{\tiny RIS}}^{L\,H}_{\overline{L}\, \overline{H}} + {\gamma_\text{\tiny RIS}}^{L\,L}_{\overline{H}\, \overline{H}} \Big) \nonumber \\
& + \Bigg[ \left( 2 \ydD \yN + \ydL - \ydH \right) \rdDNtoLbarH + \left( 2\ydD - \ydH + \ydL \yN  \right) \rdDHbartoLbarN \nonumber \\
& \hspace{0.4cm} + \left( 2\ydD + \ydL - \ydH \yN \right) \rdDLtoHN \Bigg] \ ,
\end{align}
\end{subequations}
where we follow usual notations and conventions which are elaborated in appendix \ref{app:def_con_be}. The comoving number density is denoted by $Y_X = n_X/s$ which is the ratio of number density $(n_X)$ and entropy density $(s)$ for any species $X$ and is a function of the temperature $(T)$, parameterized by the dimensionless variable $z = M_N/T$. The left hand side of these differential equations denote the change in the number density of $X$ which is defined as $\dot{Y}_X = z \ s(z) \ H(z) \ dY_X/dz$ where $X = N, \Sigma \Delta, \delta \Delta$ and $\delta(B-L)$ and $H(z)$ is the Hubble parameter. 

Note that $B_L$ and $B_H$ are the branching ratios of the triplet decaying into lepton and Higgs respectively and are given by
\begin{equation} \label{eq:BL_and_BH}
B_L = \dfrac{M_\Delta}{8 \pi \Gamma_\Delta} \text{Tr} \left( {\mathcal{Y}_\Delta}^\dagger \mathcal{Y}_\Delta \right) \ ; \quad B_H = \dfrac{\mu^2}{8 \pi M_\Delta \Gamma_\Delta} \ .
\end{equation}
The superscript `eq' denotes the equilibrium number density of any given species. The asymmetry in the abundance of SM lepton doublet $(Y_{\delta L})$ and Higgs doublet $(Y_{\delta H})$ are linearly dependent on $Y_{\delta \Delta}$ and $Y_{\delta (B-L)}$ and can be expressed in terms of $C_L$ and $C_H$ coefficient matrices given in \cite{Barbieri:1999ma, Nardi:2005hs, Abada:2006fw, Sierra:2014tqa}. 
The decay reaction densities of RHN and triplet are denoted by $\gDN$ and $\gDD$ respectively which also act as source terms (proportional to $\epsilon_N$ and $\epsilon_\Delta$) in the evolution of $B-L$ asymmetry. All other reaction densities correspond to $2 \leftrightarrow 2$ scattering processes and are schematically represented by $\gamma^X_Y \equiv \gamma(X \rightarrow Y)$. The real intermediate state (RIS) subtracted reaction densities are denoted as $\gamma_\text{\tiny RIS}$ for $\Delta L = 2$ processes as discussed later and all of the reaction densities for various processes are explicitly summarized in appendix \ref{app:reac_den}. 

In passing we note that we do not include any decay or scattering process that involves three or more particles in the initial or final state \cite{Nardi:2007jp} and also exclude gauge interactions related to pure type I seesaw which involves thermal corrections to masses of gauge bosons and other particles \cite{Giudice:2003jh}. In the following we list down various processes and the corresponding reaction densities relevant for this hybrid framework. \\

\noindent  $\bullet$ \textbf{Decays of RHN and triplet:} \\

\noindent The RHN decays into a lepton and a Higgs, whereas the triplet can decay into a lepton pair or a Higgs pair as shown in figure \ref{fig:decay_N} and \ref{fig:decay_D} respectively. Both the decay processes violate lepton number and the associated reaction densities for RHN and triplet decay are given as
\begin{equation} \label{eq:rd_decay}
\gDN = n_N^\text{eq} \dfrac{K_1 (z)}{K_2 (z)} \Gamma_N \ ; \quad \gDD = n_{\Sigma \Delta}^\text{eq} \dfrac{K_1 (z)}{K_2 (z)} \Gamma_\Delta \ ,
\end{equation}
where the common ratio of the Bessel functions accounts for the dilution arising from the expansion of the Universe \cite{Buchmuller:2004nz}. $\Gamma_N$ and $\Gamma_\Delta$ denotes the total decay width of RHN and triplet respectively and given by eq. \ref{eq:decay_widths}. \\

\noindent $\bullet$ \textbf{Scattering processes: ${\Delta L = 1}$} \\

\noindent The $\Delta L = 1$ scattering process can be broadly divided into two classes, one that is already present in the pure type I seesaw framework and the other that is only possible in the mixed type I + II seesaw scenario. There are three such processes that involve the top quark in the external legs as shown in figure \ref{fig:scat_pure_type1}. The squared amplitudes of these processes are given in appendix \ref{eq:app:scat_pure_type1}. 
\begin{figure}[h!]
	\centering
	\begin{subfigure}{0.32 \textwidth}
		\centering
		\scalebox{1}{
		\begin{tikzpicture}[baseline=(current bounding box.center)]
		\begin{feynman}
			\vertex (i12) ;
			\vertex [above left = of i12] (i1) {\large $N$} ;
			\vertex [below left = of i12] (i2) {\large $L$} ;
			\vertex [right = of i12] (f12) ;
			\vertex [above right = of f12] (f1) {\large $\overline{Q}$};
			\vertex [below right = of f12] (f2) {\large $t_R$};

			\diagram{
				(i1) -- [plain, very thick] (i12) -- [scalar, very thick, edge label' = {\large $\overline{H}$}] (f12) -- [anti fermion, very thick] (f1) ;
				(i2) -- [fermion, very thick] (i12) ;
				(f12) -- [fermion, very thick] (f2) ;
				};
		\end{feynman}
		\end{tikzpicture}
		}
	\end{subfigure} \hspace{0.5cm}
	\begin{subfigure}{0.32 \textwidth}
		\centering
		\scalebox{1}{
		\begin{tikzpicture}[baseline=(current bounding box.center)]
		\begin{feynman}
			\vertex (if1) ;
			\vertex [left = of if1] (i1) {\large $N$} ;
			\vertex [right = of if1] (f1) {\large $L$} ;
			\vertex [below = 2.75cm of if1] (if2) ;
			\vertex [left = of if2] (i2) {\large $Q$} ;
			\vertex [right = of if2] (f2) {\large $t_R$} ;

			\diagram*{
				(i1) -- [plain, very thick] (if1) -- [fermion, very thick] (f1);
				(if1) -- [scalar, very thick, edge label = {\large \ \ $H$}] (if2) ;
				(i2) -- [fermion, very thick] (if2) -- [fermion, very thick] (f2);
				};
		\end{feynman}
		\end{tikzpicture}
		}
	\end{subfigure} \hspace{-0.5cm}
	\begin{subfigure}{0.32 \textwidth}
		\centering
		\scalebox{1}{
		\begin{tikzpicture}[baseline=(current bounding box.center)]
		\begin{feynman}
			\vertex (if1) ;
			\vertex [left = of if1] (i1) {\large $N$} ;
			\vertex [right = of if1] (f1) {\large $L$} ;
			\vertex [below = 2.75cm of if1] (if2) ;
			\vertex [left = of if2] (i2) {\large $\overline{t_R}$} ;
			\vertex [right = of if2] (f2) {\large $\overline{Q}$} ;

			\diagram*{
				(i1) -- [plain, very thick] (if1) -- [fermion, very thick] (f1);
				(if1) -- [scalar, very thick, edge label = {\large \ \ $H$}] (if2) ;
				(i2) -- [anti fermion, very thick] (if2) -- [anti fermion, very thick] (f2);
				};
		\end{feynman}
		\end{tikzpicture}
		}
	\end{subfigure}
	\caption{$2 \leftrightarrow 2 $ scattering processes $(\Delta L = 1)$ involving top quark}
	\label{fig:scat_pure_type1}
\end{figure}
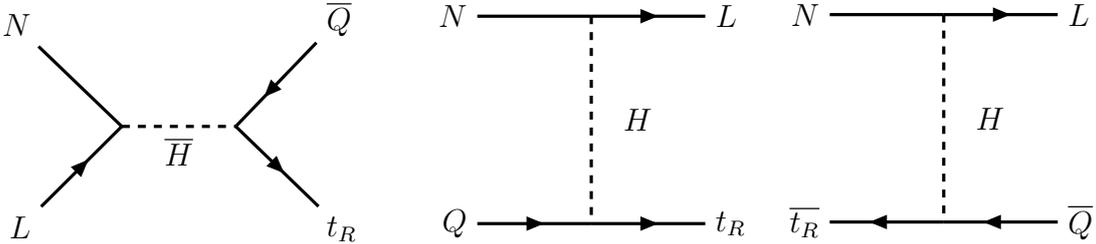

On the other hand, there are also three additional processes which involve both $N_R$ and $\Delta$ in the external legs and are only possible in this hybrid seesaw framework. They constitute the mixed topology processes that we include in the analysis for the first time. The novel processes include $\Delta N \leftrightarrow \overline{L} H$, $\Delta \overline{H} \leftrightarrow \overline{L} N$ and $\Delta L \leftrightarrow H N$ which are shown in figure \ref{fig:hybrid_DN_to_LbarH}, \ref{fig:hybrid_DHbar_to_LbarN} and \ref{fig:hybrid_DL_to_HN} with their squared amplitudes are given in appendix \ref{eq:app:scat_mixed_type1and2}. These hybrid processes in our framework are mediated either by the lepton or the Higgs. In the following section we will see that the inclusion of these processes can be quite important in certain parameter regions to significantly modify the final value of baryon asymmetry.
%
\begin{figure}[h!]
	\centering
	\begin{subfigure}{0.45\textwidth}
		\centering
		\scalebox{1}{
		\begin{tikzpicture}[baseline=(current bounding box.center)]
		\begin{feynman}
			\vertex (if1) ;
			\vertex [left = of if1] (i1) {\large $\Delta$} ;
			\vertex [right = of if1] (f1) {\large $\overline{L}$} ;
			\vertex [below = 2.75cm of if1] (if2) ;
			\vertex [left = of if2] (i2) {\large $N$} ;
			\vertex [right = of if2] (f2) {\large $H$} ;

			\diagram*{
				(i1) -- [scalar, very thick] (if1) -- [anti fermion, very thick] (f1);
				(if1) -- [fermion, very thick, edge label = {\large \ \ $L$}] (if2) ;
				(i2) -- [plain, very thick] (if2) -- [scalar, very thick] (f2);
			};

		\end{feynman}
		\end{tikzpicture}
		}
		\caption{}
		\label{fig:hybrid_DN_to_LbarH_Lmed}
	\end{subfigure}
	\begin{subfigure}{0.45\textwidth}
		\centering
		\scalebox{1}{
		\begin{tikzpicture}[baseline=(current bounding box.center)]
		\begin{feynman}
			\vertex (if1) ;
			\vertex [left = of if1] (i1) {\large $\Delta$} ;
			\vertex [right = of if1] (f1) {\large $H$} ;
			\vertex [below = 2.75cm of if1] (if2) ;
			\vertex [left = of if2] (i2) {\large $N$} ;
			\vertex [right = of if2] (f2) {\large $\overline{L}$} ;

			\diagram*{
					(i1) -- [scalar, very thick] (if1) -- [scalar, very thick] (f1);
					(if1) -- [scalar, very thick, edge label = {\large \ \ $H$}] (if2) ;
					(i2) -- [plain, very thick] (if2) -- [anti fermion, very thick] (f2);
				};
		\end{feynman}
		\end{tikzpicture}
		}
		\caption{}
		\label{fig:hybrid_DN_to_LbarH_Hmed}
	\end{subfigure}
	\caption{Lepton (left) and Higgs (right) mediated diagram for $\Delta N \leftrightarrow \overline{L} H$ scattering}
	\label{fig:hybrid_DN_to_LbarH}
\end{figure}
\begin{figure}[h!]
	\centering
	\begin{subfigure}{0.45\textwidth}
		\centering
		\scalebox{1}{
		\begin{tikzpicture}[baseline=(current bounding box.center)]
		\begin{feynman}
				\vertex (i12) ;
				\vertex [above left = of i12] (i1) {\large $\Delta$} ;
				\vertex [below left = of i12] (i2) {\large $\overline{H}$} ;
				\vertex [right = of i12] (f12) ;
				\vertex [above right = of f12] (f1) {\large $\overline{L}$};
				\vertex [below right = of f12] (f2) {\large $N$};

				\diagram{
						(i1) -- [scalar, very thick] (i12) -- [scalar, very thick, edge label' = {\large $H$}] (f12) -- [anti fermion, very thick] (f1) ;
						(i2) -- [scalar, very thick] (i12) ;
						(f12) -- [plain, very thick] (f2) ;
					};
			\end{feynman}
		\end{tikzpicture}
		}
		\caption{}
		\label{fig:hybrid_DHbar_to_LbarN_Hmed}
	\end{subfigure}
	\begin{subfigure}{0.45\textwidth}
		\centering
		\scalebox{1}{
		\begin{tikzpicture}[baseline=(current bounding box.center)]
		\begin{feynman}
			\vertex (if1) ;
			\vertex [left = of if1] (i1) {\large $\Delta$} ;
			\vertex [right = of if1] (f1) {\large $\overline{L}$} ;
			\vertex [below = 2.75cm of if1] (if2) ;
			\vertex [left = of if2] (i2) {\large $\overline{H}$} ;
			\vertex [right = of if2] (f2) {\large $N$} ;

			\diagram*{
					(i1) -- [scalar, very thick] (if1) -- [anti fermion, very thick] (f1);
					(if1) -- [fermion, very thick, edge label = {\large \ \ $L$}] (if2) ;
					(i2) -- [scalar, very thick] (if2) -- [plain, very thick] (f2);
				};
		\end{feynman}
		\end{tikzpicture}
		}
		\caption{}
		\label{fig:hybrid_DHbar_to_LbarN_Lmed}
	\end{subfigure}
	\caption{Higgs (left) and lepton (right) mediated diagram for $\Delta \overline{H} \leftrightarrow \overline{L} N$ scattering}
	\label{fig:hybrid_DHbar_to_LbarN}
\end{figure}
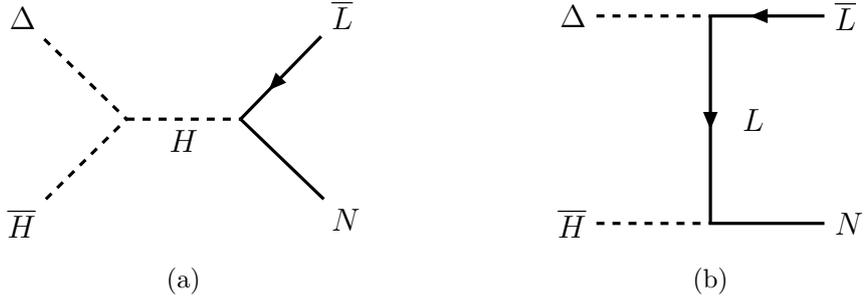
\begin{figure}[h!]
	\centering
	\begin{subfigure}{0.45\textwidth}
		\centering
		\scalebox{1}{
		\begin{tikzpicture}[baseline=(current bounding box.center)]
		\begin{feynman}
				\vertex (i12) ;
				\vertex [above left = of i12] (i1) {\large $\Delta$} ;
				\vertex [below left = of i12] (i2) {\large $L$} ;
				\vertex [right = of i12] (f12) ;
				\vertex [above right = of f12] (f1) {\large $H$};
				\vertex [below right = of f12] (f2) {\large $N$};

				\diagram{
						(i1) -- [scalar, very thick] (i12) -- [fermion, very thick, edge label' = {\large $L$}] (f12) -- [scalar, very thick] (f1) ;
						(i2) -- [fermion, very thick] (i12) ;
						(f12) -- [plain, very thick] (f2) ;
					};
			\end{feynman}
		\end{tikzpicture}
		}
		\caption{}
		\label{fig:hybrid_DL_to_HN_Lmed}
	\end{subfigure}
	\begin{subfigure}{0.45\textwidth}
		\centering
		\scalebox{1}{
		\begin{tikzpicture}[baseline=(current bounding box.center)]
		\begin{feynman}
			\vertex (if1) ;
			\vertex [left = of if1] (i1) {\large $\Delta$} ;
			\vertex [right = of if1] (f1) {\large $H$} ;
			\vertex [below = 2.75cm of if1] (if2) ;
			\vertex [left = of if2] (i2) {\large $L$} ;
			\vertex [right = of if2] (f2) {\large $N$} ;

			\diagram*{
					(i1) -- [scalar, very thick] (if1) -- [scalar, very thick] (f1);
					(if1) -- [scalar, very thick, edge label = {\large \ \ $H$}] (if2) ;
					(i2) -- [fermion, very thick] (if2) -- [plain, very thick] (f2);
				};
		\end{feynman}
		\end{tikzpicture}
		}
		\caption{}
		\label{fig:hybrid_DL_to_HN_Hmed}
	\end{subfigure}
	\caption{Lepton (left) and Higgs (right) mediated diagram for $\Delta L \leftrightarrow H N$ scattering}
	\label{fig:hybrid_DL_to_HN}
\end{figure}
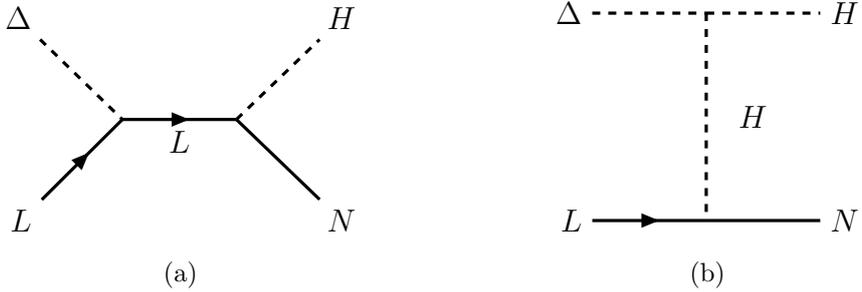
\ \\

\noindent $\bullet$ \textbf{Scattering processes: ${\Delta L = 2}$} \\

\noindent Apart from inverse decays and $\Delta L = 1$ scattering processes, one needs to include the $\Delta L = 2$ scatterings which are of the same order in the couplings as the asymmetry parameters. We consider the two processes $L H \leftrightarrow \overline{L} \, \overline{H}$ and $L L \leftrightarrow \overline{H} \, \overline{H}$ which are mediated by $N$ and $\Delta$ as shown in figure \ref{fig:LH_to_LbarHbar} and \ref{fig:LL_to_HbarHbar}. 
\begin{figure}[h!]
	\centering
	\begin{subfigure}{0.32\textwidth}
		\centering
		\scalebox{1}{
		\begin{tikzpicture}[baseline=(current bounding box.center)]
		\begin{feynman}
				\vertex (i12) ;
				\vertex [above left = of i12] (i1) {\large $L$} ;
				\vertex [below left = of i12] (i2) {\large $H$} ;
				\vertex [right = of i12] (f12) ;
				\vertex [above right = of f12] (f1) {\large $\overline{L}$};
				\vertex [below right = of f12] (f2) {\large $\overline{H}$};

				\diagram{
						(i1) -- [fermion, very thick] (i12) -- [plain, very thick, edge label' = {\large $N$}] (f12) -- [anti fermion, very thick] (f1) ;
						(i2) -- [scalar, very thick] (i12) ;
						(f12) -- [scalar, very thick] (f2) ;
					};
			\end{feynman}
		\end{tikzpicture}
		}
		\caption{}
		\label{fig:LH_to_LbarHbar_N_s}
	\end{subfigure}
	\begin{subfigure}{0.32\textwidth}
		\centering
		\scalebox{1}{
		\begin{tikzpicture}[baseline=(current bounding box.center)]
		\begin{feynman}
			\vertex (if1) ;
			\vertex [left = of if1] (i1) {\large $L$} ;
			\vertex [right = of if1] (f1) {\large $\overline{L}$} ;
			\vertex [below = 2.75cm of if1] (if2) ;
			\vertex [left = of if2] (i2) {\large $H$} ;
			\vertex [right = of if2] (f2) {\large $\overline{H}$} ;

			\diagram*{
					(i1) -- [fermion, very thick] (if1);
					(if1) -- [plain, very thick, edge label' = {\large $N$ \ \ } ] (if2) ;
					(i2) -- [scalar, very thick] (if2);
					(if1) -- [scalar, very thick] (f2);
					(if2) -- [anti fermion, very thick] (f1);
				};
		\end{feynman}
		\end{tikzpicture}
		}
		\caption{}
		\label{fig:LH_to_LbarHbar_N_u}
	\end{subfigure}
	\begin{subfigure}{0.32\textwidth}
		\centering
		\scalebox{1}{
		\begin{tikzpicture}[baseline=(current bounding box.center)]
		\begin{feynman}
			\vertex (if1) ;
			\vertex [left = of if1] (i1) {\large $L$} ;
			\vertex [right = of if1] (f1) {\large $\overline{L}$} ;
			\vertex [below = 2.75cm of if1] (if2) ;
			\vertex [left = of if2] (i2) {\large $H$} ;
			\vertex [right = of if2] (f2) {\large $\overline{H}$} ;

			\diagram*{
					(i1) -- [fermion, very thick] (if1) -- [anti fermion, very thick] (f1);
					(if1) -- [scalar, very thick, edge label = {\large \ \ $\Delta$} ] (if2) ;
					(i2) -- [scalar, very thick] (if2) -- [scalar, very thick] (f2);
				};
		\end{feynman}
		\end{tikzpicture}
		}
		\caption{}
		\label{fig:LH_to_LbarHbar_D_t}
	\end{subfigure}
	\caption{RHN mediated s-channel (left), u-channel (middle), and triplet mediated t-channel (right) Feynman diagrams contributing to the $\Delta L = 2$ scattering process $L H \leftrightarrow \overline{L} \, \overline{H}$}
	\label{fig:LH_to_LbarHbar}
\end{figure}
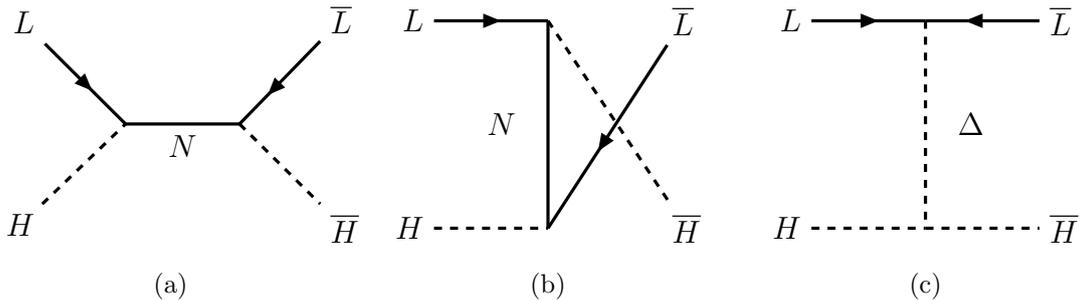
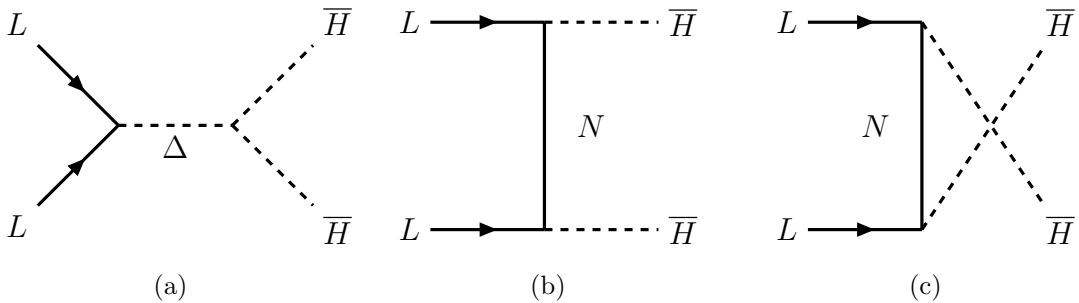
\begin{figure}[h!]
	\centering
	\begin{subfigure}{0.32\textwidth}
		\centering
		\scalebox{1}{
		\begin{tikzpicture}[baseline=(current bounding box.center)]
		\begin{feynman}
				\vertex (i12) ;
				\vertex [above left = of i12] (i1) {\large $L$} ;
				\vertex [below left = of i12] (i2) {\large $L$} ;
				\vertex [right = of i12] (f12) ;
				\vertex [above right = of f12] (f1) {\large $\overline{H}$};
				\vertex [below right = of f12] (f2) {\large $\overline{H}$};

				\diagram{
						(i1) -- [fermion, very thick] (i12) -- [scalar, very thick, edge label' = {\large $\Delta$}] (f12) -- [scalar, very thick] (f1) ;
						(i2) -- [fermion, very thick] (i12) ;
						(f12) -- [scalar, very thick] (f2) ;
					};
			\end{feynman}
		\end{tikzpicture}
		}
		\caption{}
		\label{fig:LL_to_HbarHbar_D_s}
	\end{subfigure}
	\begin{subfigure}{0.32\textwidth}
		\centering
		\scalebox{1}{
		\begin{tikzpicture}[baseline=(current bounding box.center)]
		\begin{feynman}
			\vertex (if1) ;
			\vertex [left = of if1] (i1) {\large $L$} ;
			\vertex [right = of if1] (f1) {\large $\overline{H}$} ;
			\vertex [below = 2.75cm of if1] (if2) ;
			\vertex [left = of if2] (i2) {\large $L$} ;
			\vertex [right = of if2] (f2) {\large $\overline{H}$} ;

			\diagram*{
					(i1) -- [fermion, very thick] (if1) -- [scalar, very thick] (f1);
					(if1) -- [plain, very thick, edge label = {\large \ \ $N$} ] (if2) ;
					(i2) -- [fermion, very thick] (if2) -- [scalar, very thick] (f2);
				};
		\end{feynman}
		\end{tikzpicture}
		}
		\caption{}
		\label{fig:LL_to_HbarHbar_N_t}
	\end{subfigure}
	\begin{subfigure}{0.32\textwidth}
		\centering
		\scalebox{1}{
		\begin{tikzpicture}[baseline=(current bounding box.center)]
		\begin{feynman}
			\vertex (if1) ;
			\vertex [left = of if1] (i1) {\large $L$} ;
			\vertex [right = of if1] (f1) {\large $\overline{H}$} ;
			\vertex [below = 2.75cm of if1] (if2) ;
			\vertex [left = of if2] (i2) {\large $L$} ;
			\vertex [right = of if2] (f2) {\large $\overline{H}$} ;

			\diagram*{
					(i1) -- [fermion, very thick] (if1);
					(if1) -- [plain, very thick, edge label' = {\large $N$ \ \ } ] (if2) ;
					(i2) -- [fermion, very thick] (if2);
					(if1) -- [scalar, very thick] (f2);
					(if2) -- [scalar, very thick] (f1);
				};
		\end{feynman}
		\end{tikzpicture}
		}
		\caption{}
		\label{fig:LL_to_HbarHbar_N_u}
	\end{subfigure}
	\caption{Triplet mediated s-channel (left), RHN mediated t-channel (middle) and u-channel (right) Feynman diagrams contributing to the $\Delta L = 2$ scattering process $L L \leftrightarrow \overline{H} \, \overline{H}$}
	\label{fig:LL_to_HbarHbar}
\end{figure}
It is important to note that in order to avoid double counting, one must consider only the off-shell contribution by subtraction of RIS \cite{Kolb:1979qa, Davidson:2008bu} given by 
\begin{equation}
{\gamma_\text{\tiny RIS}}^X_Y = \gamma^X_Y - {\gamma_\text{\tiny on-shell}}^X_Y = \gamma^X_Y - \gamma^X_U \ \text{BR}(U \rightarrow Y) \ ,
\end{equation}
where $U$ is the intermediate unstable particle. The relevant expressions can be found in appendix \ref{eq:app:deltal2}.

\section{Matter antimatter asymmetry in full and approximated analysis} \label{sec:num_analysis_results}

In this section we present the estimated present-day baryon asymmetry obtained by performing a numerical solution of the coupled differential equations given in eq. \ref{eq:boltz_full} for the hybrid framework depicted in section \ref{sec:model}. We solve the Boltzmann equations tracking the asymmetry created through a synergy of $N$ and $\Delta$ decays keeping all mixed processes discussed above. This will reflect the ``full analysis" in the discussions that follows. We compare this analysis with hierarchical scenarios where the leptogenesis is dominated with asymmetries either created by the decay of RHN $(N)$ or triplet $(\Delta)$. In this approximation the mixed processes shown in figures \ref{fig:hybrid_DN_to_LbarH}, \ref{fig:hybrid_DHbar_to_LbarN} and \ref{fig:hybrid_DL_to_HN} are not present. Henceforth, we refer to the scenario where leptogenesis is dominated by the decay of RHN as ``$N$-approximate" and the situation where the leptogenesis is dominated by the decay of scalar triplet as ``$\Delta$-approximate" for convenience. The BEs for these two approximate situations can be obtained by systematically switching off certain terms from eq. \ref{eq:boltz_full} as sketched above.
\begin{table}[h!]
\centering \setlength\arraycolsep{5pt} \Large 
\renewcommand{\arraystretch}{1.5}
\resizebox{\textwidth}{!}{
\begin{tabular}{|c|c|c|c|c|c|c|c|c|c|c|c|c|c|c|c|c|c|c|}
\hline 
\multicolumn{19}{|c|}{Model parameters for the benchmark points}\tabularnewline
\hline 
\hline 
 & $M_N$ [GeV] & $M_\Delta$ [GeV] & \multicolumn{5}{c|}{$\mathcal{Y}_\nu / 10^{-2}$} & \multicolumn{10}{c|}{$\mathcal{Y}_\Delta / 10^{-2}$} & $\mu$ [GeV] \tabularnewline
\hline 
BP1 & $1.038 \times 10^{14}$ & $2.154 \times 10^{12}$ & \multicolumn{5}{c|}{$\begin{pmatrix}         
		2.24337 + i \  2.52128 \\         
		4.91356 + i \  2.62518 \\         
		1.80870 + i \  3.17213     
	\end{pmatrix}$} & \multicolumn{10}{c|}{$\begin{pmatrix}
    		2.10519 & 0.94246 + i \ 2.10688 & 2.55296 + i \ 0.94298 \\
        \cdot & 10.83470 & 10.77910 + i \ 2.55374 \\
        \cdot & \cdot & 7.91460
    \end{pmatrix}$} & $1.881 \times 10^{10}$ \tabularnewline
\hline 
BP2 & $9.714 \times 10^{13}$ & $2.878 \times 10^{12}$ & \multicolumn{5}{c|}{$\begin{pmatrix}         
		0.24558 + i \  0.77045 \\         
		0.54368 + i \  0.07879 \\         
		0.27885 + i \  0.14405     
	\end{pmatrix}$} & \multicolumn{10}{c|}{$\begin{pmatrix}
    		0.14271 & 0.07116 + i \ 0.14434 & 0.24941 + i \ 0.07004 \\
        \cdot & 1.04974 & 0.92076 + i \ 0.25063 \\
        \cdot & \cdot & 0.73111
    \end{pmatrix}$} & $3.689 \times 10^{11}$ \tabularnewline
\hline 
\hline 
\multicolumn{19}{|c|}{Neutrino oscillation parameters within $3\sigma$ in normal hierarchy (F)} \tabularnewline
\hline 
\hline 
 & \multicolumn{2}{|c|}{$\Delta m_{12}^2$ [eV$^2$]} & \multicolumn{2}{|c|}{$\Delta m_{13}^2$ [eV$^2$]} & \multicolumn{3}{|c|}{$\theta_{12}$} & \multicolumn{3}{|c|}{$\theta_{23}$} & \multicolumn{3}{|c|}{$\theta_{13}$} & \multicolumn{2}{|c|}{$\delta_\text{CP}$} & \multicolumn{1}{|c|}{$J_\text{CP}$} & \multicolumn{1}{|c|}{$\sum m_\nu$ [eV]} & \multicolumn{1}{|c|}{$|m_{\beta \beta}|$ [eV]} \tabularnewline 
\hline 
BP1 & \multicolumn{2}{|c|}{$7.4081 \times 10^{-5}$} & \multicolumn{2}{|c|}{$2.5745 \times 10^{-3}$} & \multicolumn{3}{|c|}{34.0376\degree} & \multicolumn{3}{|c|}{48.1313\degree} & \multicolumn{3}{|c|}{8.7770\degree} & \multicolumn{2}{|c|}{153.078\degree} & \multicolumn{1}{|c|}{-0.019640} & \multicolumn{1}{|c|}{0.06427} & \multicolumn{1}{|c|}{0.0067} \tabularnewline 
\hline 
BP2 & \multicolumn{2}{|c|}{$7.4427 \times 10^{-5}$} & \multicolumn{2}{|c|}{$2.5620 \times 10^{-3}$} & \multicolumn{3}{|c|}{31.4090\degree} & \multicolumn{3}{|c|}{49.3888\degree} & \multicolumn{3}{|c|}{8.3471\degree} & \multicolumn{2}{|c|}{157.071\degree} & \multicolumn{1}{|c|}{-0.017706} & \multicolumn{1}{|c|}{0.06274} & \multicolumn{1}{|c|}{0.0056} \tabularnewline 
\hline 
\hline
\multicolumn{19}{|c|}{Leptogenesis related quantities} \tabularnewline
\hline 
\hline 
 & \multicolumn{3}{|c|}{$\epsilon_N$} & \multicolumn{3}{|c|}{$\epsilon_D$} & \multicolumn{3}{|c|}{$\epsilon_r$} & \multicolumn{1}{c|}{$Y_{\delta B}^0 (F)$} & \multicolumn{6}{c|}{$Y_{\delta B}^0 (N)$} & \multicolumn{2}{c|}{$Y_{\delta B}^0 (\Delta)$} \tabularnewline
\hline 
BP1 & \multicolumn{3}{|c|}{$-4.704 \times 10^{-6}$} & \multicolumn{3}{|c|}{$-1.442 \times 10^{-7}$} & \multicolumn{3}{|c|}{$32.6$}  & \multicolumn{1}{c|}{$8.772 \times 10^{-11}$}  & \multicolumn{6}{c|}{$17.548 \times 10^{-11}$} & \multicolumn{2}{c|}{$6.896 \times 10^{-11}$} \tabularnewline
\hline 
BP2 & \multicolumn{3}{|c|}{$-2.788 \times 10^{-6}$} & \multicolumn{3}{|c|}{$-4.398 \times 10^{-9}$} & \multicolumn{3}{|c|}{$633.8$} & \multicolumn{1}{c|}{$8.803 \times 10^{-11}$}  & \multicolumn{6}{c|}{$3.149 \times 10^{-9}$} & \multicolumn{2}{c|}{$1.234 \times 10^{-12}$} \tabularnewline
\hline
\end{tabular}
\renewcommand{\arraystretch}{1.0}
}
\caption{Model parameters for the two benchmark points which satisfy neutrino oscillation data in normal hierarchy as well as reproduce the correct baryon asymmetry in the full analyses denoted by $(F)$. The final value of $Y_{\delta B}$ in the $N$-approximate and $\Delta$-approximate scenarios are denoted by $(N)$ and $(\Delta)$ respectively along with the numerical values of CP asymmetry parameter. The superscript $(0)$ indicates the final value of the baryon asymmetry in different scenarios.}
\label{tab:BPs}
\end{table}

For a quantitative comparison between the full analysis of the hybrid framework with the approximate scenarios specially where there exists a relative closeness between the two seesaw states we present a numerical simulation of two benchmark points given in table \ref{tab:BPs}. For the hybrid seesaw framework we can envisage three different choices of satisfying the neutrino oscillation data. (I) Neutrino observables are solely generated by the contribution of $N_R$ which is not feasible in our minimal hybrid setup. (II) Owing to large number of independent free Yukawa couplings, the triplet alone can generate all the neutrino oscillation observables within experimental uncertainties with negligible contribution from RHN. This scenario has been depicted in our second benchmark point (BP2). (III) Finally, the neutrino oscillation data can be generated by contributions of both type I and type II seesaw mechanisms which has been shown in our first benchmark point (BP1). The two benchmark points BP1 and BP2 have been identified to represent the two extreme possibilities within the minimal type I + II hybrid seesaw framework that can be in consonance with neutrino oscillation data. For each benchmark point we perform a detailed study of the neutrino oscillation parameters. We investigate the relative contribution of the two seesaw mechanisms to the neutrino mass matrix. Next, we study the generation of baryon asymmetry through leptogenesis for these benchmark points in the full analysis and in the $N/\Delta$-approximate scenario. 

\begin{figure}[t!]
	\centering
	\begin{subfigure}{0.485\textwidth}
		\centering
		\includegraphics[width=1\linewidth]{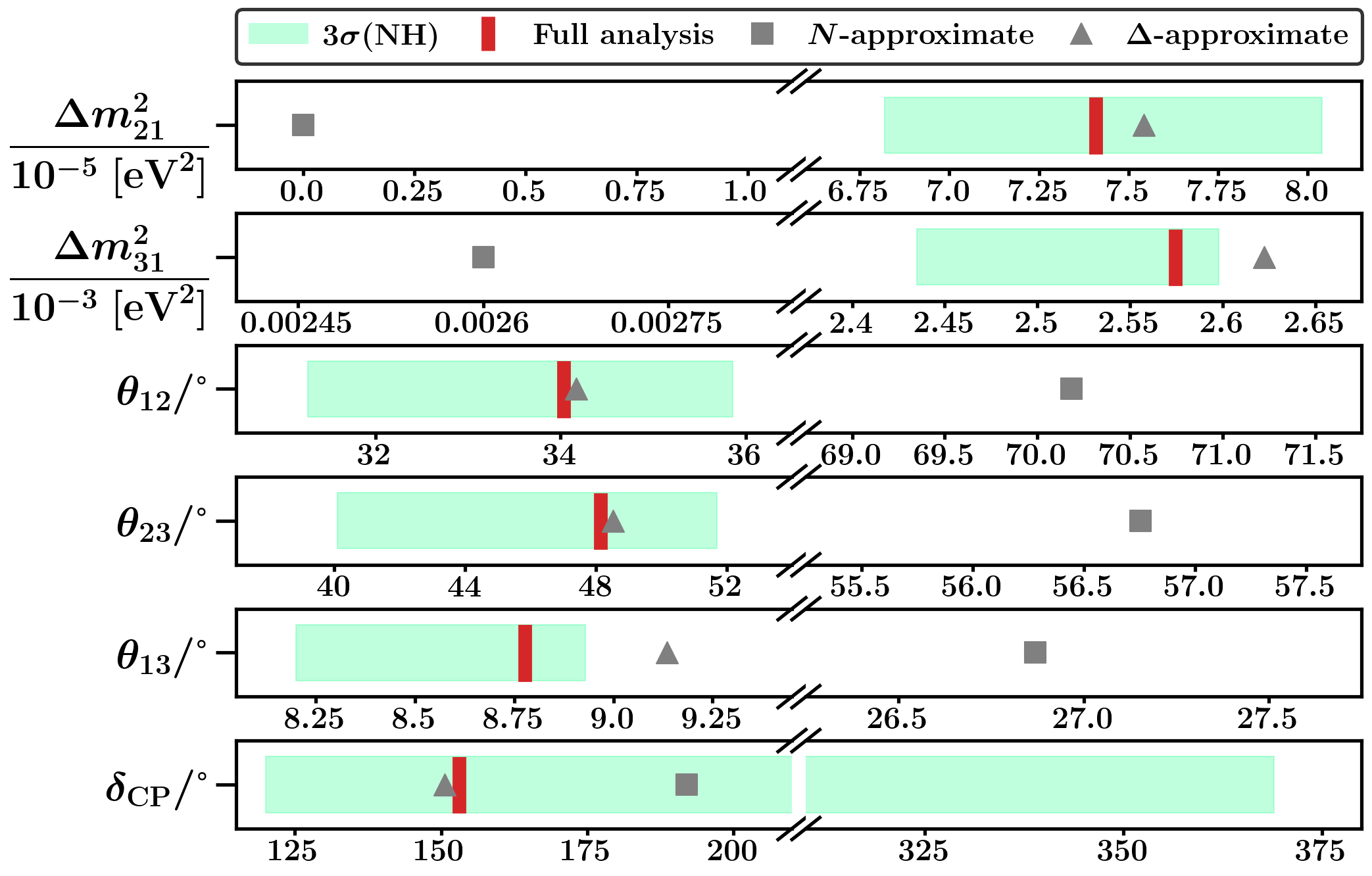} 
		\caption{}
		\label{fig:BPfinal1_nuosc_comparison}
	\end{subfigure} 
	\begin{subfigure}{0.505\textwidth}
		\centering
		\includegraphics[width=1\linewidth]{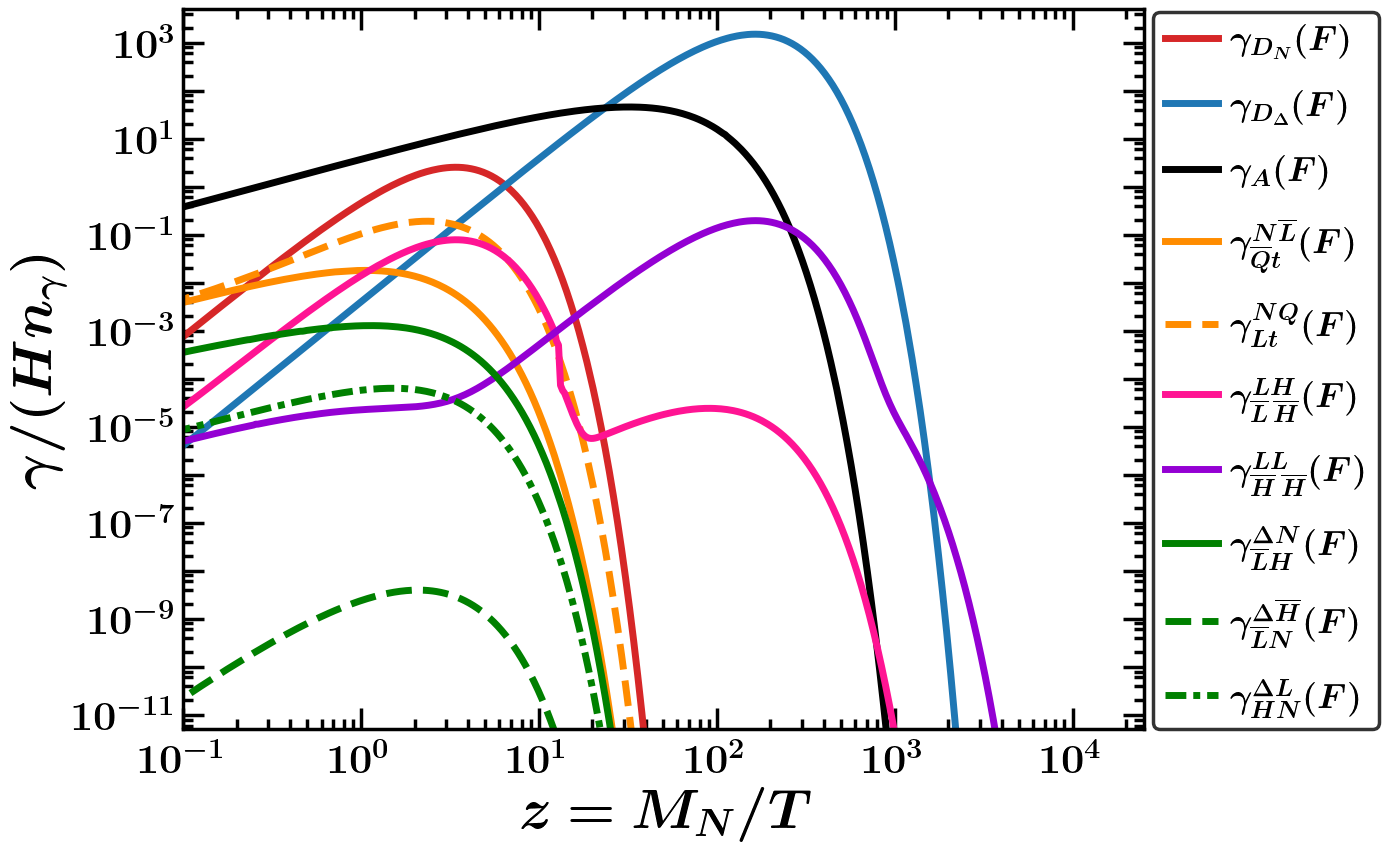} 
		\caption{}
		\label{fig:BPfinal1_reac_densities}
	\end{subfigure}
	\caption{BP1: Left panel (\ref{fig:BPfinal1_nuosc_comparison}) denotes the relative contribution of $N$ and $\Delta$ via type I and II seesaw mechanisms respectively with $3\sigma$ allowed oscillation parameters \cite{Esteban:2020cvm} for normal hierarchy (NH). The red vertical line gives the values of the oscillation parameters for the full analysis whereas the gray square and triangle are the values obtained by only taking into the contribution of type I and type II seesaw respectively. Right panel (\ref{fig:BPfinal1_reac_densities}) shows the reaction densities (scaled by the product of Hubble parameter and photon number density) for different processes involved in the hybrid seesaw scenario with respect to $z = M_N/T$ which are calculated in full analysis denoted by $(F)$. }
	\label{fig:BPfinal1_nuosc_and_reac}
\end{figure}
\begin{figure}[t!]
	\centering
	\begin{subfigure}{0.4815\textwidth}
		\centering
		\includegraphics[width=1\linewidth]{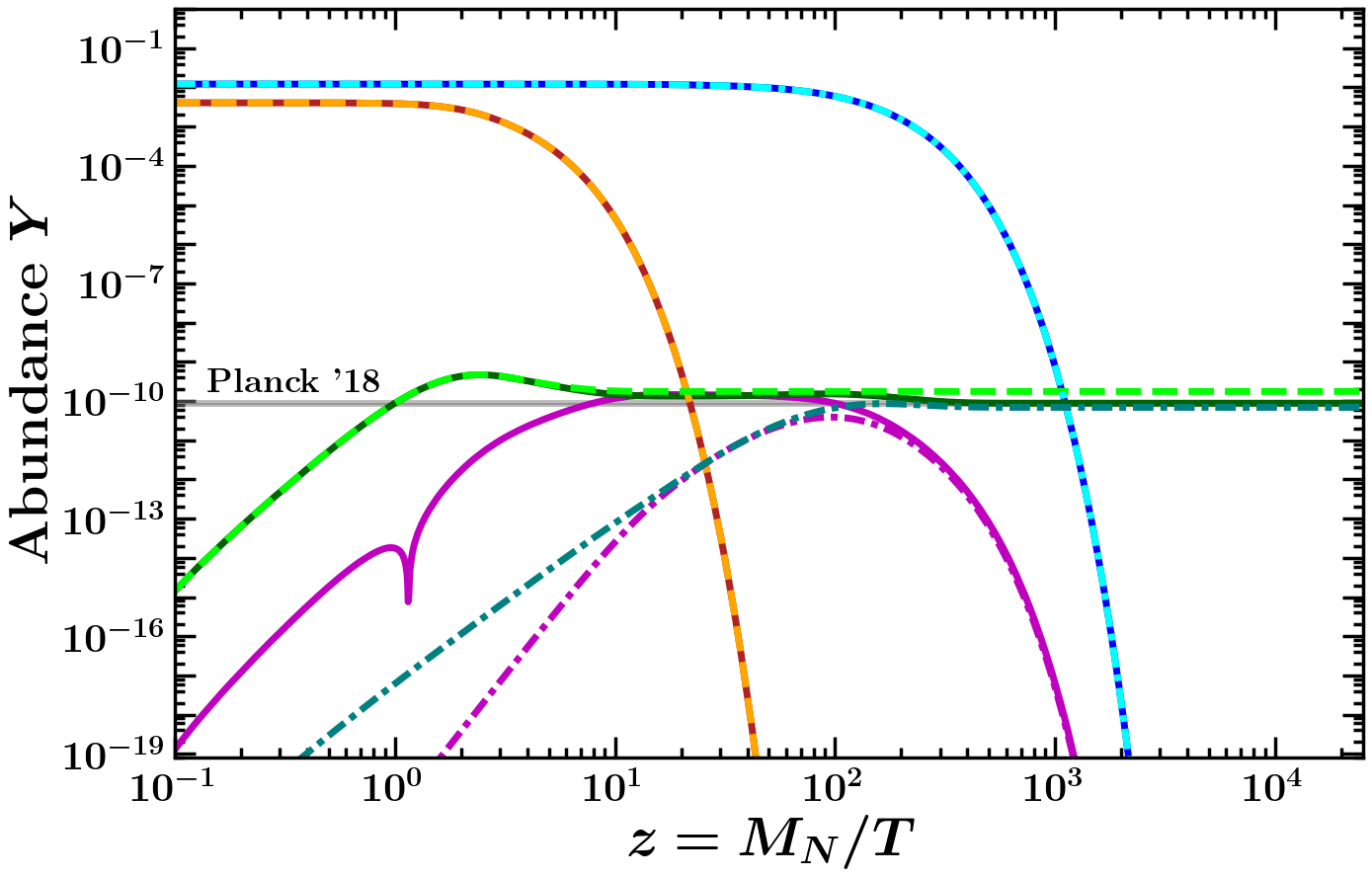} 
		\caption{}
		\label{fig:BPfinal1_abun}
	\end{subfigure} 
	\begin{subfigure}{0.5085\textwidth}
		\centering
		\includegraphics[width=1\linewidth]{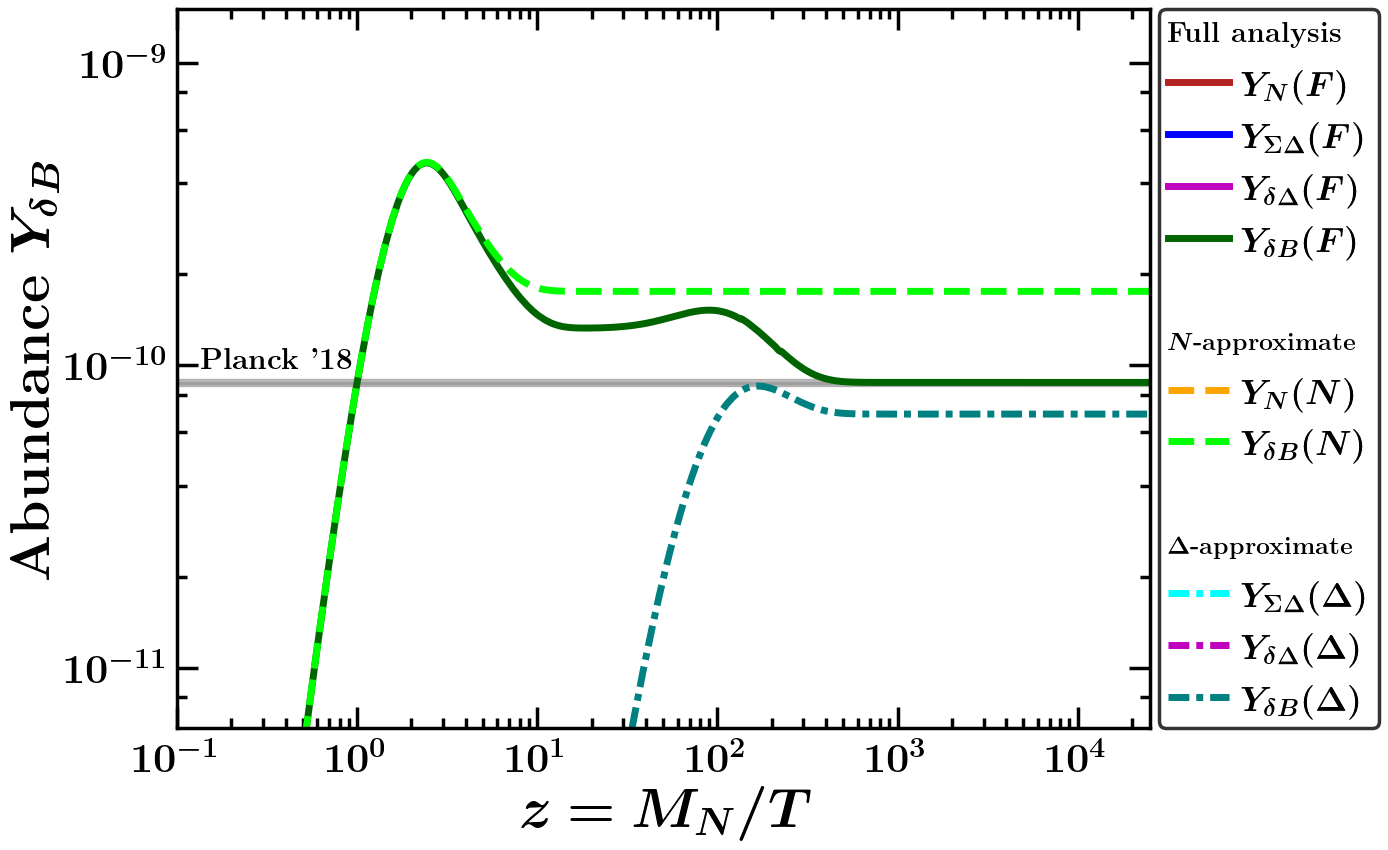} 
		\caption{}
		\label{fig:BPfinal1_abun_Y_B}
	\end{subfigure}
	\caption{BP1: Left panel (\ref{fig:BPfinal1_abun}) demonstrates the evolution of various abundances for the scenario of full analysis, $N$-approximate and $\Delta$-approximate indicated by solid, dashed and dot dashed lines and denoted by $F, N$ and $\Delta$ in the parentheses respectively. Right panel (\ref{fig:BPfinal1_abun_Y_B}) shows only the evolution of the baryon asymmetry for these three scenarios.}
	\label{fig:BPfinal1_abundance_both}
\end{figure}

In the case of the first benchmark point BP1 given in table \ref{tab:BPs}, the triplet contribution to the neutrino mass is dominant over the RHN, however the $\Delta m_{13}^2$ and $\theta_{13}$ satisfies the $3\sigma$ allowed ranges for normal hierarchy (NH) only when both the type I and II contributions are taken into account as shown in figure \ref{fig:BPfinal1_nuosc_comparison}. The reaction densities for all the processes calculated in the full analysis denoted by $(F)$ are shown in figure \ref{fig:BPfinal1_reac_densities}. The evolution of various abundances are plotted against $z = M_N/T$ in figure \ref{fig:BPfinal1_abun} for the full analysis, $N$-approximate and $\Delta$-approximate analysis scenarios denoted by solid $(F)$, dashed $(N)$ and dotted $(\Delta)$ lines respectively. The abundances of RHN $(Y_N)$ and triplet $(Y_{\Sigma\Delta})$ in the $N$-approximate and $\Delta$-approximate case follows the full analysis whereas the abundance of asymmetric component of the triplet $Y_{\delta \Delta}$ in the $\Delta$-approximate scenario differs significantly from the full analysis. Owing to higher initial value of $\epsilon_N$ the baryon asymmetry calculated in the full analysis initially follow the $N$-approximate result and starts deviating around $z \sim 5$. Interestingly the asymmetry shows a peak like behavior at $z \sim 100$ as the triplet goes on-shell in the $L L \rightarrow \overline{H} \, \overline{H}$ process. The baryon asymmetry calculated in the full analysis starts to saturate around $z \sim 500$ and matches with the observed value measured by Planck \cite{Planck:2018vyg}. As indicated in figure \ref{fig:BPfinal1_abun_Y_B} the $N$-approximate ($\Delta$-approximate) scenario exhibit a deviation of the order of $\sim 100 \, (20) \%$ from the full analysis.
\begin{figure}[t!]
	\centering
	\begin{subfigure}{0.485\textwidth}
		\centering
		\includegraphics[width=1\linewidth]{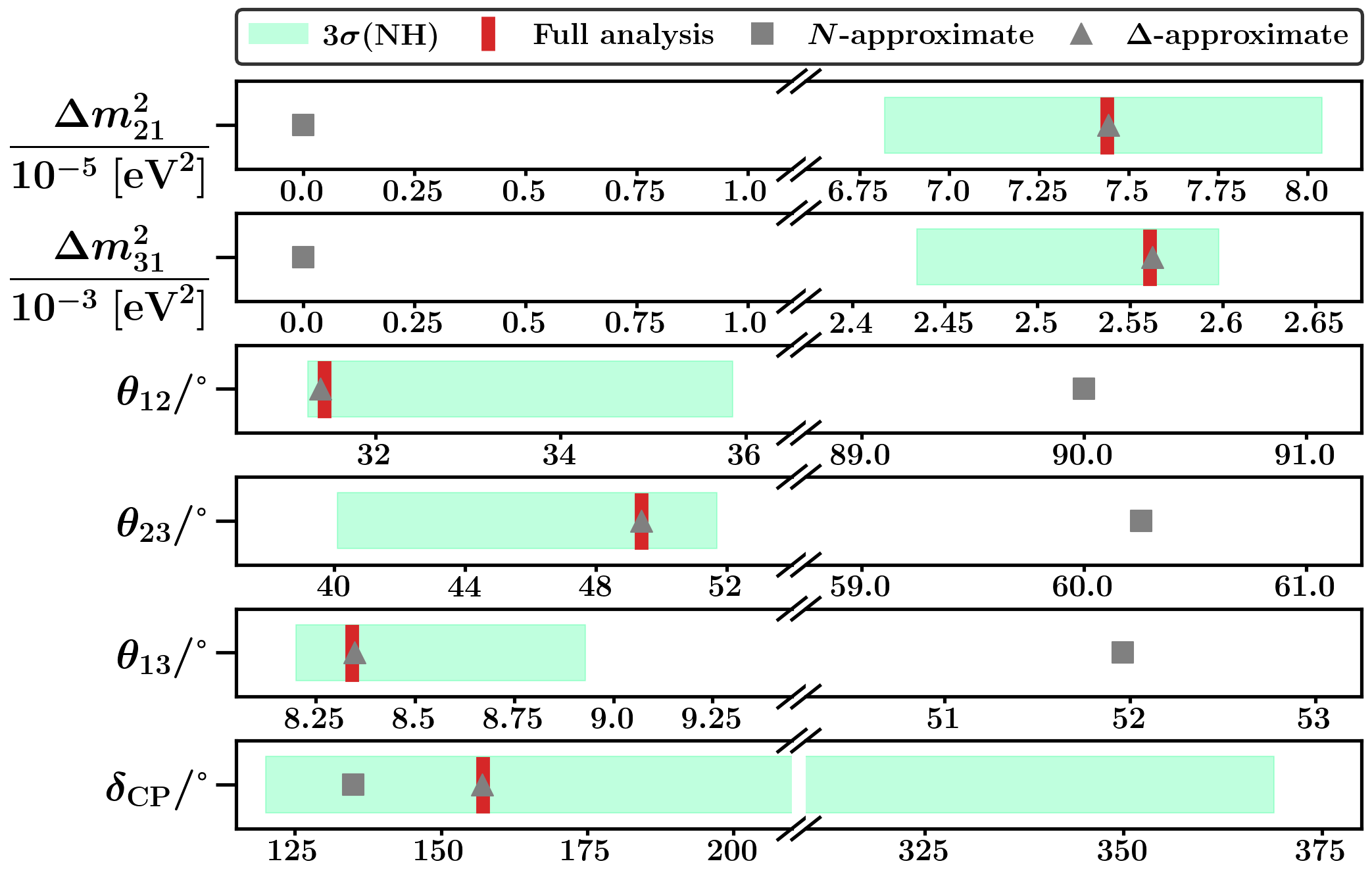} 
		\caption{}
		\label{fig:BPfinal2_nuosc_comparison}
	\end{subfigure} 
	\begin{subfigure}{0.505\textwidth}
		\centering
		\includegraphics[width=1\linewidth]{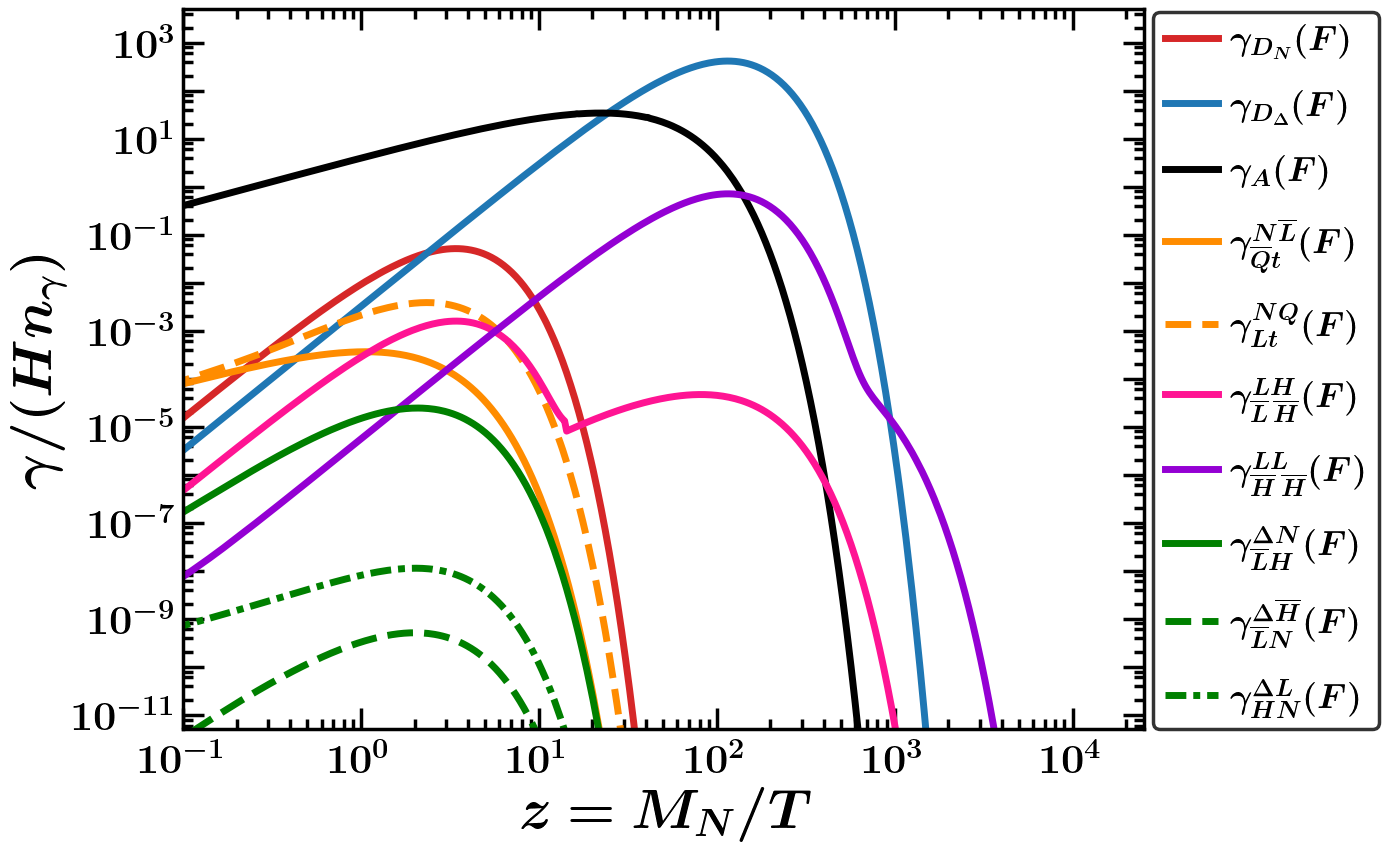} 
		\caption{}
		\label{fig:BPfinal2_reac_densities}
	\end{subfigure}
	\caption{BP2: Same as of figure \ref{fig:BPfinal1_nuosc_and_reac}. }
	\label{fig:BPfinal2_nuosc_and_reac}
\end{figure}
\begin{figure}[t!]
	\centering
	\begin{subfigure}{0.4815\textwidth}
		\centering
		\includegraphics[width=1\linewidth]{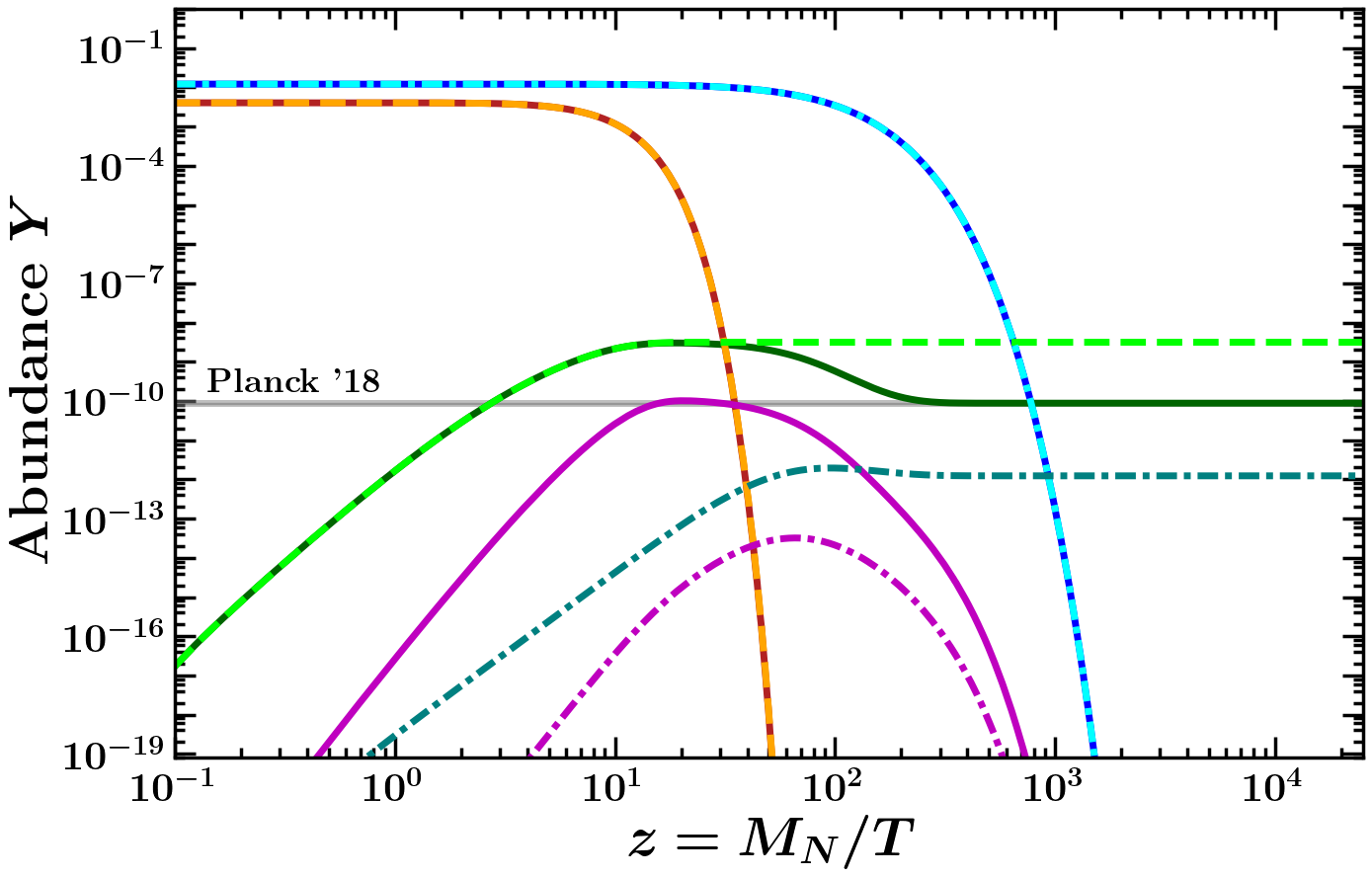} 
		\caption{}
		\label{fig:BPfinal2_abun}
	\end{subfigure} 
	\begin{subfigure}{0.5085\textwidth}
		\centering
		\includegraphics[width=1\linewidth]{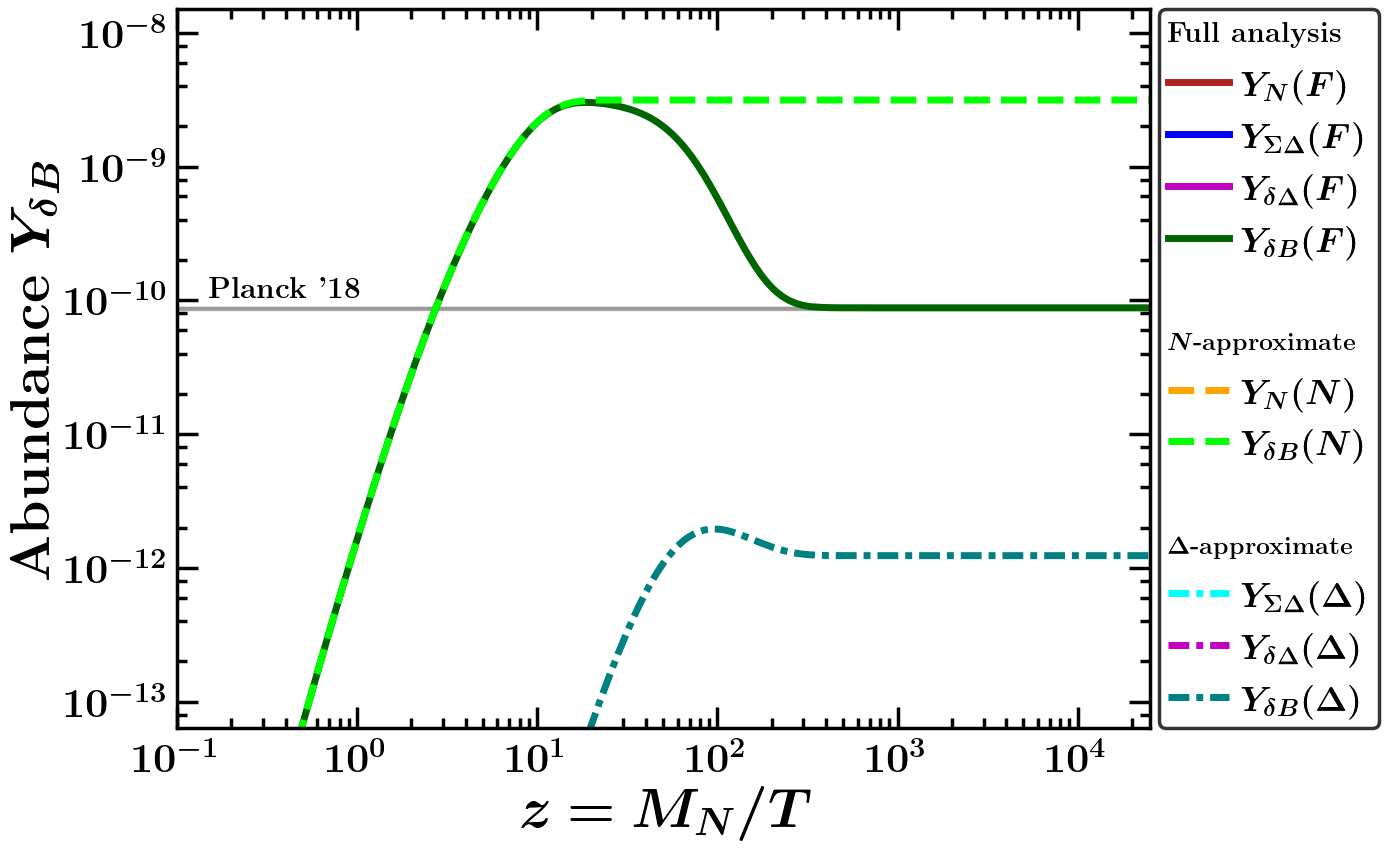} 
		\caption{}
		\label{fig:BPfinal2_abun_Y_B}
	\end{subfigure}
	\caption{BP2: Same as of figure \ref{fig:BPfinal1_abundance_both}. }
	\label{fig:BPfinal2_abundance_both}
\end{figure}

A similar behavior of the baryon asymmetry can be observed for the second benchmark point BP2 (as defined in table \ref{tab:BPs}), although all the oscillation parameters are solely determined by the type II contributions, as can be read off from figure \ref{fig:BPfinal2_nuosc_comparison}. Due to a high value of $\epsilon_r$, the baryon asymmetry is controlled initially by the RHN upto $z \sim 10$ as can be seen from figure \ref{fig:BPfinal2_abun_Y_B}. For this benchmark point we obtain a factor $100$ deviation in the final baryon asymmetry for both the approximated scenarios from the full analysis.

These two representative benchmark points indicate that in spite of the lower scalar triplet mass the baryon asymmetry is not well estimated by the $\Delta$-approximate scenario. This is indicative of the phenomena that with relatively close mass scales the underlying mixed processes play a significant role in determining the final baryon asymmetry. Thus a full analysis keeping all mixed and $\Delta L = 2$ processes revealing the full synergy of the two seesaws is crucial for an accurate determination of the baryon asymmetry in hybrid scenarios.

\section{Validity of approximate analysis in hybrid leptogenesis} \label{sec:validity_of_approximation}

We present a quantitative analysis for the region of validity of the approximate analysis of leptogenesis with hierarchical hybrid scenarios. We consider a representative RHN mass at $M_N = 9.714 \times 10^3$ GeV and fix the Yukawa couplings at $\mathcal{Y}_\nu \sim \mathcal{O}(10^{-2})$ and $\mathcal{Y}_\Delta \sim \mathcal{O}(10^{-3})$. The exact values chosen are given in table \ref{tab:BPmassvary} for completeness. The trilinear coupling to triplet mass ratio is also kept unchanged at $\mu/M_\Delta = 0.038$ and the ratio of CP asymmetry parameter is $\epsilon_r = 1.227$. 
\begin{table}[h!]
\centering \setlength\arraycolsep{5pt}
\renewcommand{\arraystretch}{1.25}
\begin{tabular}{|c|c|}
\hline
$\mathcal{Y}_\nu / 10^{-2}$ & $\mathcal{Y}_\Delta / 10^{-3}$ \tabularnewline
\hline
\hline
$\begin{pmatrix}         
		2.456 + i \  7.704 \\         
		5.437 + i \  0.788 \\         
		2.788 + i \  1.440     
	\end{pmatrix}$ & $\begin{pmatrix}
    		5.708 & 2.846 + i \ 5.774 & 9.976 + i \ 2.802 \\
        \cdot & 41.990 & 36.830 + i \ 10.025 \\
        \cdot & \cdot & 29.244
    \end{pmatrix}$ \tabularnewline
\hline
\end{tabular}
\renewcommand{\arraystretch}{1.0}
\caption{Yukawa couplings of RHN and the scalar triplet}
\label{tab:BPmassvary}
\end{table}

\begin{figure}[h!]
	\centering
	\begin{subfigure}{0.49\textwidth}
		\centering
		\includegraphics[width=1\linewidth]{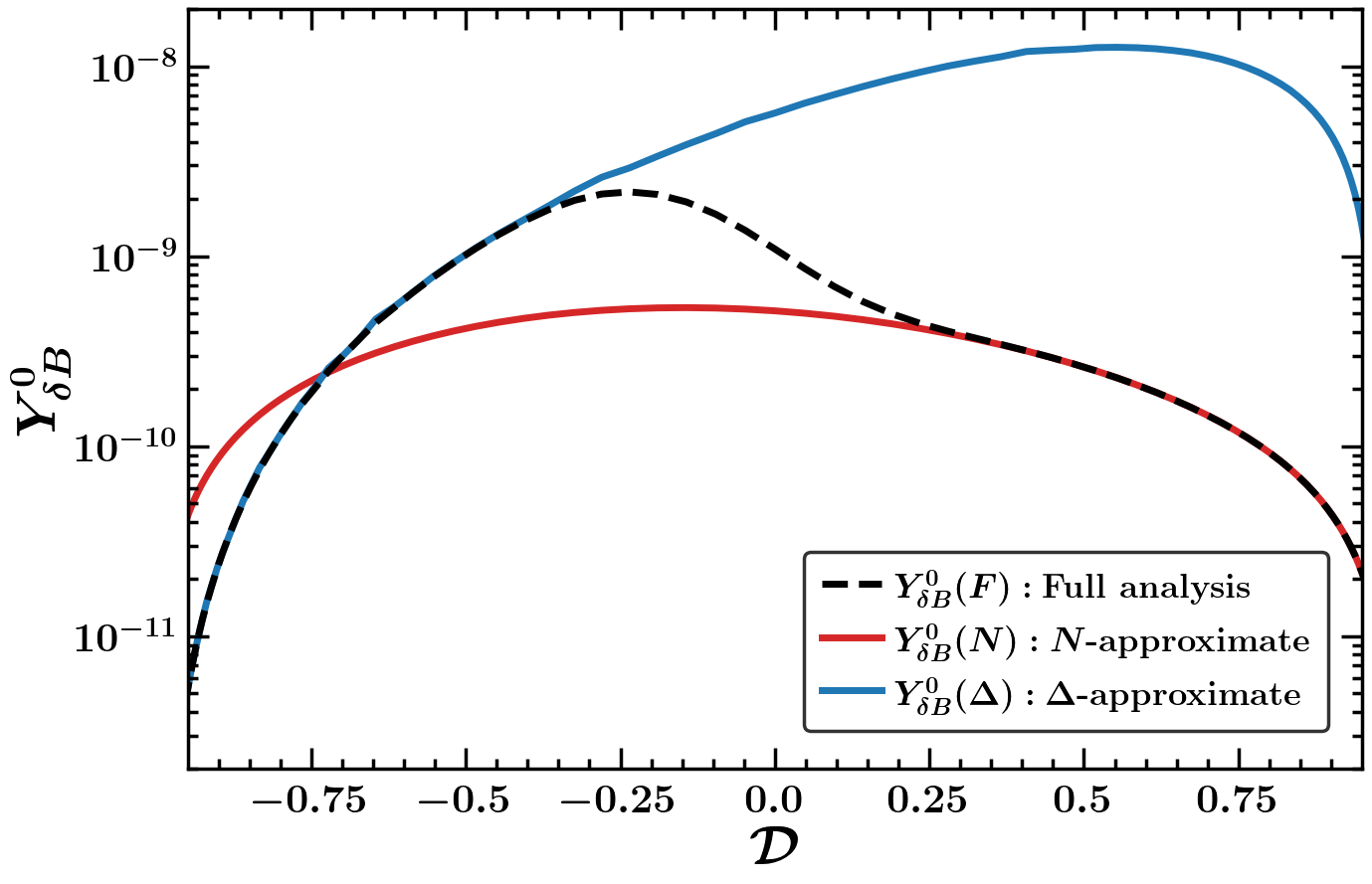} 
		\caption{}
		\label{fig:BPmassvary_degen_param_vs_final_value}
	\end{subfigure} 
	\begin{subfigure}{0.4925\textwidth}
		\centering
		\includegraphics[width=1\linewidth]{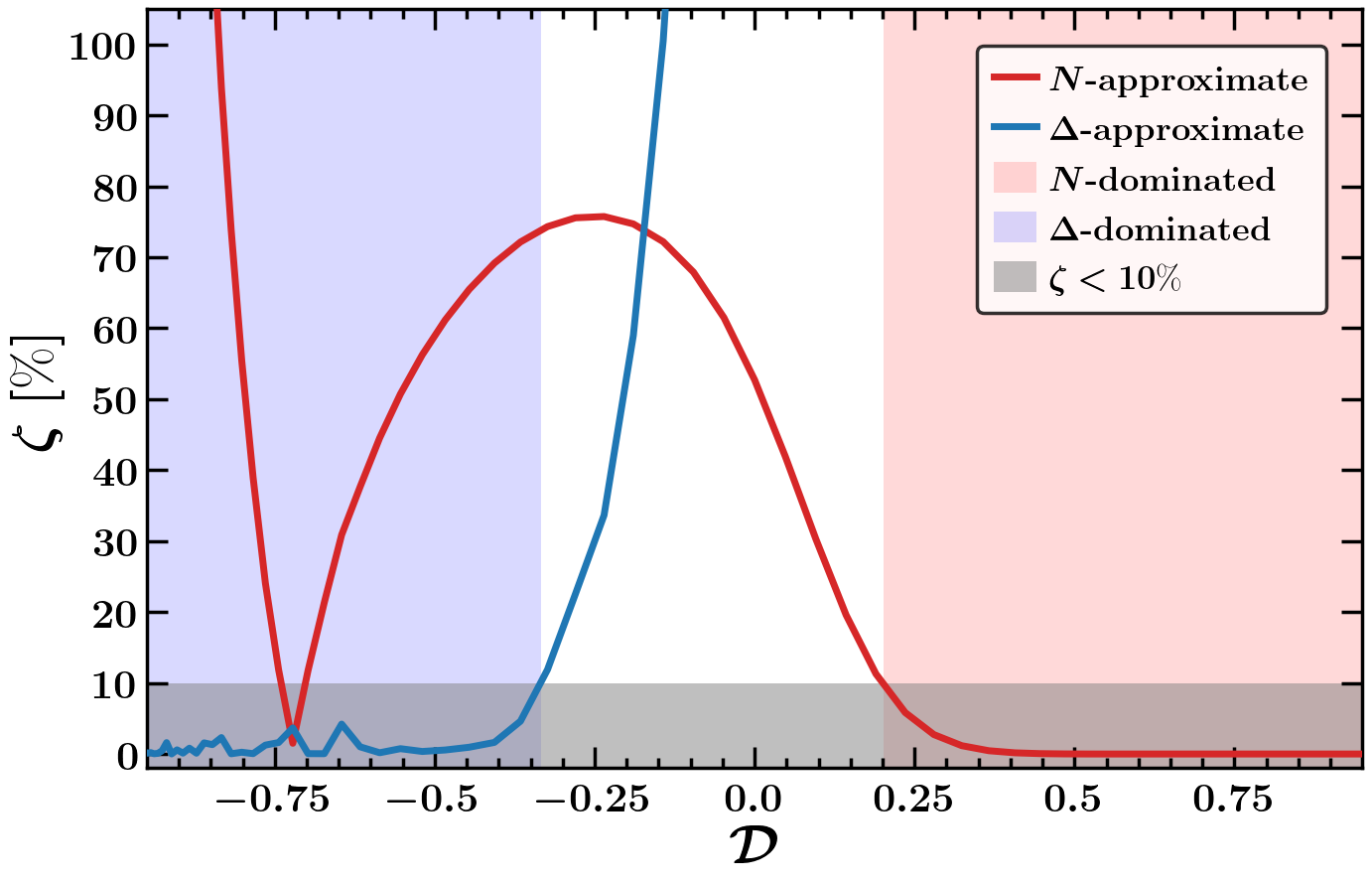} 
		\caption{}
		\label{fig:BPmassvary_degen_param_vs_error}
	\end{subfigure}
	\caption{Left panel shows the final value of baryon asymmetry in the full analysis, $N$-approximate and $\Delta$-approximate scenarios denoted by black dashed, solid red and solid blue lines respectively. Right panel denotes the error which one obtains in the $N (\Delta)$-approximate calculations in comparison with the full analysis denoted by blue (red) solid curves. The red (blue) shaded region denotes that the $N (\Delta)$-approximate calculations are in well approximations of the full analysis with an error of less than $10 \%$ denoted by the gray shaded region.}
	\label{fig:BPfinal2_abundance_both}
\end{figure}

We define a degeneracy parameter $\mathcal{D}$ given by
\begin{equation} \label{eq:degen_param}
\mathcal{D} \equiv \dfrac{r -1}{r + 1} = \dfrac{M_\Delta - M_N}{M_\Delta + M_N} \ ,
\end{equation}
which ranges between $[-1, \ 1]$ and gives an estimate of the mass degeneracy in the hybrid leptogenesis scenario and the error with respect to the full analysis is defined as
\begin{equation}
\zeta = \dfrac{Y_{\delta B}^0 - Y_{\delta B}^0(F)}{Y_{\delta B}^0(F)} \ .
\end{equation}
The final value of the baryon asymmetry is calculated in the full analysis and in approximated scenarios. This is plotted against the degeneracy parameter shown in figure \ref{fig:BPmassvary_degen_param_vs_final_value}. The relative error $\zeta$ in the $N/\Delta$-approximate scenarios are plotted as a function of degeneracy parameter in figure \ref{fig:BPmassvary_degen_param_vs_error}.
The lightest of the decaying particle usually dictates the fate of the final baryon asymmetry in the case of hierarchical spectrum with comparable values of CP asymmetry parameters. As can be clearly seen in figure \ref{fig:BPmassvary_degen_param_vs_final_value} for large degeneracy with $\mathcal{D} \gtrsim |0.25|$ the full analysis represented by the black dashed line agrees with the $\Delta (N)$-approximate results represented by the blue (red) solid lines. More quantitatively this region maps to the blue (red) shaded region in figure \ref{fig:BPmassvary_degen_param_vs_error} where the $\Delta (N)$-approximate analysis is within $10 \%$ of the results obtained with the full calculation.

As the seesaw masses come closer, the mixed topology processes become numerically significant. The approximate result starts deviating from the complete analysis modulo local numerical artifacts arising due to fine tuned cancellations. Given the competing errors from thermal corrections \cite{Giudice:2003jh} at $10 \%$ we can consider $\mathcal{D} \gtrsim |0.25|$ as the limit for the validity of the approximated results below which a more careful complete analysis including all mixed processes is warranted. In case of hierarchical primordial CP asymmetries the region of validity of the approximate solution shrinks further making the usage of the complete analysis more imperative.


\section{Conclusion}\label{sec: conclusion}

It is conceivable that under various circumstances, more than one seesaw framework is operative, simultaneously contributing to the neutrino mass parameters and driving baryogenesis through leptogenesis. The conventional approach has been to consider leptogenesis being dominated by the lightest species with the understanding that the asymmetry created at a higher scale is expected to be washed out due to the dynamics of the lighter degrees of freedom. It has been pointed out previously that this assumption gets modified when the initial asymmetry created by the CP violating decays of the heavy state(s) is substantial such that it may still account for the present-day matter-antimatter asymmetry due to incomplete washout. 

In this work we point out a complementary scenario in a hybrid seesaw framework where the interplay of both the heavier and lighter species remains important. As the mass scales approach each other, certain (often neglected) scattering processes involving both the states become numerically significant. Even with moderate hierarchy of scales, these mixed topology processes require a complete tracing of the asymmetry keeping all the species in play. The final asymmetry in this case can significantly differ from the usual approximated estimates obtained by assuming leptogenesis being dominated by the lightest seesaw state.

We demonstrate the impact of such mixed processes for a hybrid type I + II leptogenesis framework. We show that for certain regions of the parameter space the complete analysis including these novel processes can result in more than $100 \%$ correction in the final asymmetry as compared to the approximated results. The region of validity of the approximate result crucially depends on the extent of degeneracy between the two seesaw scales. While we demonstrate the importance of the complete analysis with the inclusion of the mixed topology scattering processes within the context of a specific scenario, the implications are more general and would be applicable to any hybrid leptogenesis framework.

\acknowledgments We thank Avirup Shaw, Deep Ghosh and Debajit Bose for discussions. RP acknowledges MHRD, Government of India for the research fellowship. AS acknowledges the support from grants CRG/2021/005080 and MTR/2021/000774 from SERB, Govt. of India. The authors also acknowledge the computational support provided from the Department of Physics, IIT Kharagpur. The authors also acknowledge the National Supercomputing Mission (NSM) for providing computing resources of `PARAM Shakti' at IIT Kharagpur, which is implemented by C-DAC and supported by the Ministry of Electronics and Information Technology (MeitY) and Department of Science and Technology (DST), Government of India.

\begin{appendix}

\renewcommand{\thesection}{\Alph{section}}
\renewcommand{\theequation}{\thesection-\arabic{equation}} 

\setcounter{equation}{0} 

\section{Notation and conventions utilised in the Boltzmann equations} \label{app:def_con_be}

The evolution of the number density $n_X$ for a species $X$ is conventionally described by the Boltzmann transport equation given by,
\begin{align} \label{eq:app:generic_be}
\dfrac{dY_X}{dz} &= - \dfrac{1}{z s(z) H(z)} \mathlarger{\mathlarger{\sum}}_{i_j, f_k}  \left[ \dfrac{Y_X}{Y_X^\text{eq}} \cdot \prod_{j = 1}^{m} \dfrac{Y_{i_j}}{Y_{i_j}^\text{eq}} \cdot \gamma\left( X + \sum_{j = 1}^{m} i_j \longrightarrow \sum_{k = 1	}^{n} f_k \right) \right. \nonumber \\ 
& \hspace{4.15cm} - \left.  \prod_{k = 1}^{n} \dfrac{Y_{f_k}}{Y_{f_k}^\text{eq}} \cdot \gamma\left(\sum_{k = 1}^{n} f_k \longrightarrow X + \sum_{j = 1}^{m} i_j \right) \right] \ ,
\end{align}
where $Y_X = n_X/s$ is the comoving number density and $z = M_N/T$. The Hubble parameter $H(z)$ and the entropy density $s(z)$ is given by
\begin{equation} \label{eq:app:H_and_s}
H(z) = \sqrt{\dfrac{8 g_*}{\pi}} \dfrac{M_N^2}{M_\text{Pl}} \dfrac{1}{z^2} \ ; \quad s(z) = \dfrac{2 \pi^2}{45} g_* \left( \dfrac{M_N}{z} \right)^2 \ ,
\end{equation}
where we calculate the effective degrees of freedom in the thermal bath is given by $g_* = 114.5$ and the Planck mass is set at $M_\text{Pl} = 1.22 \times 10^{19}$ GeV. The equilibrium number density of RHN, triplet and SM (lepton and Higgs) doublets appearing in eq. \ref{eq:boltz_full} are respectively given by
\begin{equation} \label{eq:app:num_den_eq}
n_N^\text{eq} = \dfrac{1}{\pi^2} M_N^3 \dfrac{K_2 (z)}{z}   ; \quad n_{\Sigma \Delta}^\text{eq} = \dfrac{3r^2}{\pi^2} M_\Delta^3 \dfrac{K_2 (rz)}{z}  ; \quad n_{L/H}^\text{eq} = \dfrac{2}{\pi^2} \left( \dfrac{M_N}{z} \right)^3 \ ,
\end{equation}
where $r$ is the mass ratio of triplet to the RHN defined above as $r = M_\Delta / M_N$ throughout the text.

\section{Calculation of reaction densities} \label{app:reac_den}

In this section we provide a systematic calculation of all the reaction densities used in this work. The reaction density of any generic process $X + i_1 + i_2 + i_3 + \cdots i_m \longrightarrow f_1 + f_2 + f_3 + \cdots f_n$ is given by
\begin{align} \label{eq:app:reac_den_general}
\gamma^{X \, i_1 \, i_2 \, i_3 \, \cdots i_m}_{f_1 \, f_2 \, f_3 \, \cdots f_n} &\equiv \gamma \left( X + \sum_{j = 1}^{m} i_j \rightarrow \sum_{k = 1	}^{n} f_k \right) \nonumber \\
&= \int \dfrac{d^3 p_X}{ (2 \pi)^3 \, 2 E_X} f_X(E_X) \prod_{j = 1}^{m} \dfrac{d^3 p_{i_j}}{ (2 \pi)^3 \, 2 E_{i_j}} f_{i_j}(E_{i_j}) \prod_{k = 1}^{n} \dfrac{d^3 p_{f_k}}{ (2 \pi)^3 \, 2 E_{f_k}} \Big[1 \pm f_{f_k} (E_{f_k}) \Big] \nonumber \\
&\times (2 \pi)^4 \delta \left( p_X + \sum_{j = 1}^{m} p_{i_j} - \sum_{k = 1}^{n} p_{f_k} \right) \times \left| \mathcal{M} \left( X + \sum_{j = 1}^{m} i_j \rightarrow \sum_{k = 1	}^{n} f_k \right) \right|^2 \ ,
\end{align}
where $f_p(E_p) \sim e^{- E_p/T}$  is the distribution function of any particle $p$ as a function of the its energy $(E_p)$ and $|\mathcal{M}|^2$ is the squared transition amplitude of that specific process. The positive or negative sign in the factor $\big[1 \pm f_{f_k} (E_{f_k}) \big]$ depends on whether the $k$-th final state is a boson or a fermion, however, in the dilute gas approximation one can approximate the factor to unity. We also use Maxwell-Boltzmann distribution function for the initial state particles for various processes.

For $1 \rightarrow 2$ decay the general formula given in eq. \ref{eq:app:reac_den_general} simplifies to \cite{Davidson:2008bu}
\begin{equation}
\gamma^X_{f_1 \, f_2} = n_X^\text{eq} \dfrac{K_1(z)}{K_2(z)} \Gamma_X \ ,
\end{equation}
where $\Gamma_X$ is the decay width of $X$ in the rest frame.

For a generic $2 \rightarrow 2$ scattering process the reaction densities in the c.o.m frame is given by \cite{Giudice:2003jh, Davidson:2008bu}
\begin{equation} \label{eq:app:rd_wrt_s}
\gamma^{1 \, 2}_{3 \, 4} \equiv \gamma (1 \, 2 \rightarrow 3 \, 4) = \dfrac{T}{64 \pi^4}\int_{s_\text{min}}^{\infty} ds \, \sqrt{s} \, K_1 (z \sqrt{s}) \, \widehat{\sigma} (s) \ ,
\end{equation}
with $s_\text{min} = \text{Max} \left[ \, (m_1 + m_2)^2, (m_3 + m_4)^2 \, \right]$  and the reduced cross section $\widehat{\sigma}$ given by \cite{Luty:1992un}
\begin{equation} \label{eq:app:sig_hat_t_int}
\widehat{\sigma}(s) = \dfrac{1}{8 \pi s} \int_{t_-}^{t_+} \big|\mathcal{M}_{1 \, 2 \rightarrow 3 \, 4} \, (s, t) \big|^2 \ dt \ ,
\end{equation}
where $t$ is the Mandelstam variable with the limits \cite{ParticleDataGroup:2022pth}
\begin{equation} \label{eq:app:t_limits}
t_\pm = \dfrac{1}{4s} \left[ \left(m_1^2 - m_2^2 - m_3^2 + m_4^2 \right)^2 - \left\lbrace \lambda^{1/2} \left(s, m_1^2, m_2^2 \right) \mp \lambda^{1/2} \left(s, m_3^2, m_4^2 \right) \right\rbrace^2  \right] \ .
\end{equation}
For a s-channel process this can be expressed in the closed form by
\begin{equation} \label{eq:app:sig_hat_s}
\widehat{\sigma} (s) = \dfrac{1}{8 \pi} \lambda^{1/2} (1, m_1^2/s, m_2^2/s) \ \lambda^{1/2} (1, m_1^2/s, m_2^2/s) \ \big|\mathcal{M}_{1 \, 2 \rightarrow 3 \, 4} \big|^2 \ , 
\end{equation}
where $\lambda$ is the Kallen function defined as $\lambda(a, b, c) \equiv a^2 + b^2 + c^2 - 2ab - 2bc - 2ca$.

It is evident from eq. \ref{eq:app:rd_wrt_s} and eq. \ref{eq:app:sig_hat_s} that the reduced cross section is dimensionless and thus it is more convenient to express several expressions in terms of dimensionless quantities defined as
\begin{equation}
x = \dfrac{s}{M_N^2} \ ; \ y = \dfrac{t}{M_N^2} \ ; \ y_\pm = \dfrac{t_\pm}{M_N^2} \ ; \ r = \dfrac{M_\Delta}{M_N} \ ; \ a_i = \dfrac{m_i^2}{M_N^2} \ ; \ a_{\Gamma_i} = \dfrac{m_i^2}{M_N^2} \ \forall \  i \in [L, H] \ ,
\end{equation}
where $a_H$ and $a_L$ are set to be $10^{-5}$ and $10^{-8}$ respectively \cite{Luty:1992un, Hahn-Woernle:2009jyb} to handle the infrared divergences. Using these dimensionless quantities the reaction density given in eq. \ref{eq:app:rd_wrt_s} and reduced cross section given in eq. \ref{eq:app:sig_hat_t_int} can be respectively written as
\begin{align}
\gamma^{1 \, 2}_{3 \, 4} = \dfrac{M_N^4}{64 \pi^4 z}\int_{x_\text{min}}^{\infty} dx \, \sqrt{x} \, K_1 (z \sqrt{x}) \, \widehat{\sigma} (x) \ ; \quad \widehat{\sigma}(x) = \dfrac{1}{8 \pi x} \int_{y_-}^{y_+} \big|\mathcal{M}_{1 \, 2 \rightarrow 3 \, 4} \, (y) \big|^2 \ dy \ ,
\end{align}
where $x_\text{min} = s_\text{min}/M_N^2$. With these notations we list down the reduced cross section of all the various processes that are important for our analysis in the following subsections.

\subsection{Standard processes within type I/II leptogenesis} \label{eq:app:scat_pure_type1}
The $\Delta L = 1$ processes that are present in the pure type I seesaw are shown in figure \ref{fig:scat_pure_type1} which involves the top quark and its Yukawa coupling with the Higgs denoted by $y_t = \sqrt{2} m_t/v_H$. \\

\noindent
$\bullet$ \textbf{Process:} $N \, (p_1) + L \, (p_2) \rightarrow \overline{Q} \, (p_3) + t_R \, (p_4)$
\begin{equation}
\begin{aligned}
x_\text{min} &\simeq 1 \ , \\
\left| \mathcal{M} \right|^2(x) &\simeq 3 \cdot 2 \, \text{Tr} \left( {\mathcal{Y}_N}^\dagger \mathcal{Y}_N \right) y_t^2 \, \left( \dfrac{x - 1}{x} \right)^2 \ , \\  
\widehat{\sigma}(x) &= \dfrac{3}{4 \pi} \text{Tr} \left( {\mathcal{Y}_N}^\dagger \mathcal{Y}_N \right) y_t^2 \left( \dfrac{x - 1}{x} \right)^2 \ . 
\end{aligned}
\end{equation}
$\bullet$ \textbf{Process:} $N \, (p_1) + \overline{Q} \, (p_2) \rightarrow L \, (p_3) + t_R \, (p_4)$ and $N \, (p_1) + t_R \, (p_2) \rightarrow L \, (p_3) + \overline{Q} \, (p_4)$
\begin{equation}
\begin{aligned}
x_\text{min} &\simeq 1 \ , \\
y_- &\simeq 1 - x \text{ and } y_+ \simeq 0 \ , \\
\left| \mathcal{M} \right|^2(y) &\simeq 3 \cdot 2 \, \text{Tr} \left( {\mathcal{Y}_N}^\dagger \mathcal{Y}_N \right) y_t^2 \, \dfrac{y(y - 1)}{(y - a_H)^2} \ , \\
\widehat{\sigma}(x) &= \dfrac{3}{4 \pi} \text{Tr} \left( {\mathcal{Y}_N}^\dagger \mathcal{Y}_N \right) y_t^2 \left( \dfrac{x - 1}{x} \right) \left\lbrace \dfrac{x - 2 + 2 a_H}{x - 1 + a_H} + \dfrac{1 - 2 a_H}{x - 1} \, \text{ln} \left( \dfrac{x - 1 + a_H}{a_H} \right) \right\rbrace \ .
\end{aligned}
\end{equation}
The reduced cross section for gauge induced triplet scattering processes in type II seesaw framework are given as \cite{Sierra:2014tqa}
\begin{equation}
\begin{aligned}
x_\text{min} &\simeq 4 r^2 \ , \\
\widehat{\sigma}(x) &= \dfrac{1}{36 \pi} \bigg[ (5 C_2 - 11 C_1) w^3 + 3(w^2 - 1) \left\lbrace 2 C_1 + C_2(w^2 - 1) \right\rbrace \, \text{ln} \left( \dfrac{1 + w}{1 - w} \right) \\
& \hspace{1.15cm} + \, (15 C_1 - 3 C_2) w \bigg] + \left( \dfrac{41 g_1^4 + 50 g_2^4}{48 \pi} \right) r^{3/2} \ , \\
& \text{with } C_1 = 3 g_1^4 + 12 g_1^2 g_2^2 + 12 g_2^4 \ ; \ C_2 = 3 g_1^4 + 12 g_1^2 g_2^2 + 6 g_2^4 \ ; \ w = \sqrt{1 - \dfrac{4r^2}{x}} \ , 
\end{aligned}
\end{equation}
where $g_1$ and $g_2$ are the gauge couplings for $U(1)$ and $SU(2)_L$ gauge groups respectively.

\subsection{New hybrid processes} \label{eq:app:scat_mixed_type1and2}

Within the hybrid framework there are three mixed scattering processes where both $N$ and $\Delta$ appear in the external legs. A discussion about their reduced cross sections are now in order. \\

\noindent
$\bullet$ \textbf{Process:} $\Delta \, (p_1) + N \, (p_2) \rightarrow \overline{L} \, (p_3) + H \, (p_4)$ 

As shown in figure \ref{fig:hybrid_DN_to_LbarH} that consists of two Feynman diagrams with amplitude denoted as $\mathcal{M}_L \, (\mathcal{M}_H)$ for lepton (Higgs) mediated t-channel processes. The total squared amplitude and the reduced cross sections are given as
\begin{align}
|\mathcal{M}|^2(y) &= \left| i\mathcal{M}_L + i\mathcal{M}_H \right|^2 = \left| \mathcal{M}_L \right|^2 + \left( \mathcal{M}_L \mathcal{M}_H^\dagger + \mathcal{M}_L^\dagger \mathcal{M}_H \right) + \left| \mathcal{M}_H \right|^2 \ , \\
\widehat{\sigma}(x) &= \int_{y_-}^{y^+} |\mathcal{M}|^2(y) dy  = \widehat{\sigma}_{LL}(x) + \widehat{\sigma}_{LH}(x) + \widehat{\sigma}_{HH}(x) \ ,
\end{align}
where
\begin{align} \label{eq:app:hybrid_DN_to_LbarH} 
x_\text{min} &= \text{Max}\left[ (r + 1)^2, (\sqrt{a_L} + \sqrt{a_H})^2 \right] = (r + 1)^2 \ , \\
y_{\pm} &= \left[ \left(r^2 - 1 - a_L + a_H \right)^2 - x^2 \left\lbrace \lambda^{1/2} \left(1, r^2/x, 1/x \right) \mp \lambda^{1/2} \left(1, a_L^2/x, a_H^2/x \right) \right\rbrace^2  \right] \ , \\
\widehat{\sigma}_{LL}(x) &= \dfrac{\text{Tr} \left( \mathcal{Y}_\Delta \mathcal{Y}_\nu \mathcal{Y}_\nu^\dagger \mathcal{Y}_\Delta^\dagger \right) }{8 \pi x} \left[ (4 a_L - x + 1) \text{ln} |y - a_L| + \dfrac{ (4 a_L - r^2)(-1 + a_H - a_L) }{y - a_L} \right]_{y_-}^{y_+}  \ , \\
\widehat{\sigma}_{LH}(x) &= \dfrac{1}{8 \pi x} \, 2\mathfrak{Re} \left[ \dfrac{\mu}{M_N} \text{Tr} \left( \mathcal{Y}_\Delta \mathcal{Y}_\nu \mathcal{Y}_\nu^\dagger \right) \right] \, \dfrac{1}{x - r^2 - 1} \ \times  \nonumber \\
& \hspace{1cm} \Bigg[ (4 a_L - r^2) \text{ln} |y - a_L| + (x - 1 - 4a_L) \text{ln} |y + x - r^2 - 1 - a_L| \Bigg]_{y_-}^{y_+} \ , \\
\widehat{\sigma}_{HH}(x) &= \dfrac{1}{8 \pi x} \, \left( \dfrac{\mu}{M_N} \right)^2 \text{Tr} \left( \mathcal{Y}_\nu \mathcal{Y}_\nu^\dagger \right) \Bigg[ \text{ln} |y + x - r^2 - 1 - a_L| - \dfrac{1 + a_L - a_H}{y + x - r^2 - 1 - a_L} \Bigg]_{y_-}^{y_+}  \ .
\end{align}

\noindent 
$\bullet$ \textbf{Process:} $\Delta \, (p_1) + \overline{H} \, (p_2) \rightarrow \overline{L} \, (p_3) + N \, (p_4)$ 

The process includes two Feynman diagrams, s-channel Higgs mediated diagram and t-channel lepton mediated diagram as shown in figure \ref{fig:hybrid_DHbar_to_LbarN} with transition amplitudes denoted by $\mathcal{M}_s$ and $\mathcal{M}_t$ respectively. The total squared amplitude and reduced cross section can be written as
\begin{align} \label{eq:app:DHbar_to_LbarN_modMsq_sighat}
|\mathcal{M}|^2(y) &= \left| i\mathcal{M}_s + i\mathcal{M}_t \right|^2 = \left| \mathcal{M}_s \right|^2 + \left( \mathcal{M}_s \mathcal{M}_t^\dagger + \mathcal{M}_s^\dagger \mathcal{M}_t \right) + \left| \mathcal{M}_t \right|^2 \ , \\
\widehat{\sigma}(x) &= \int_{y_-}^{y^+} |\mathcal{M}|^2(y) dy  = \widehat{\sigma}_{ss}(x) + \widehat{\sigma}_{st}(x) + \widehat{\sigma}_{tt}(x) \ ,
\end{align}
where 
\begin{align} \label{eq:app:hybrid_DHbar_to_LbarN}
x_\text{min} &= \text{Max}\left[ (r + \sqrt{a_H})^2, (\sqrt{a_L} + 1)^2 \right] \ , \\
y_{\pm} &= \left[ \left(r^2 - a_H - a_L + 1 \right)^2 - x^2 \left\lbrace \lambda^{1/2} \left(1, r^2/x, a_H/x \right) \mp \lambda^{1/2} \left(1, a_L^2/x, 1/x \right) \right\rbrace^2  \right] \ , \\
\widehat{\sigma}_{ss}(x) &= \dfrac{1}{8 \pi} \, \left( \dfrac{\mu}{M_N} \right)^2 \text{Tr} \left( \mathcal{Y}_\nu \mathcal{Y}_\nu^\dagger \right) \lambda^{1/2}\left( 1, r^2/x, a_H/x \right) \, \lambda^{1/2}\left( 1, a_L/x, 1/x \right) \, \dfrac{x - 1 - a_L}{(x - a_H)^2} \ , \\
\widehat{\sigma}_{st}(x) &= \dfrac{1}{8 \pi x} \, 2\mathfrak{Re} \left[ \dfrac{\mu}{M_N} \text{Tr} \left( \mathcal{Y}_\Delta \mathcal{Y}_\nu \mathcal{Y}_\nu^\dagger \right) \right] \, \dfrac{1}{x - a_H} \Bigg[ y + (4 a_L - r^2) \text{ln} |y - a_L| \Bigg]_{y_-}^{y_+}  \ , \\
\widehat{\sigma}_{tt}(x) &= - \dfrac{\text{Tr} \left( \mathcal{Y}_\Delta \mathcal{Y}_\nu \mathcal{Y}_\nu^\dagger \mathcal{Y}_\Delta^\dagger \right) }{8 \pi x} \, \Bigg[ y + \dfrac{(r^2 - 4 a_L)(1 + a_L - a_H)}{y - a_L} \nonumber \\
&\hspace{4cm} + (x - a_H - r^2 + 4 a_L) \text{ln} |y - a_L| \Bigg]_{y_-}^{y_+} \ . 
\end{align}

\noindent 
$\bullet$ \textbf{Process:} $\Delta \, (p_1) + L \, (p_2) \rightarrow H \, (p_3) + N \, (p_4)$ 

The process is depicted in figure \ref{fig:hybrid_DL_to_HN} with two Feynman diagrams mediated by s-channel lepton (as shown in figure \ref{fig:hybrid_DL_to_HN_Lmed}) and t-channel Higgs (as shown in figure \ref{fig:hybrid_DL_to_HN_Hmed}) with amplitude $\mathcal{M}_s$ and $\mathcal{M}_t$ respectively. The corresponding expressions are given by,
\begin{align} \label{eq:app:hybrid_DL_to_HN}
x_\text{min} &= \text{Max}\left[ (r + \sqrt{a_L})^2, (\sqrt{a_H} + 1)^2 \right] \ , \\
y_{\pm} &= \left[ \left(r^2 - a_L - a_H + 1 \right)^2 - x^2 \left\lbrace \lambda^{1/2} \left(1, r^2/x, a_L/x \right) \mp \lambda^{1/2} \left(1, a_H^2/x, 1/x \right) \right\rbrace^2  \right] \ , \\
\widehat{\sigma}_{ss}(x) &= \dfrac{\text{Tr} \left( \mathcal{Y}_\Delta \mathcal{Y}_\nu \mathcal{Y}_\nu^\dagger \mathcal{Y}_\Delta^\dagger \right) }{8 \pi x} \, \dfrac{1}{2 (x - a_L)^2} \, \Bigg[ y(x - r^2 + 3 a_L) \left(\dfrac{y}{2} + x - a_L - a_H \right) \nonumber \\
& \hspace{7cm} + y (r^2 - 4 a_L) (\dfrac{y}{2} - a_L - 1) \Bigg]_{y_-}^{y_+} \ , \\
\widehat{\sigma}_{st}(x) &= \dfrac{1}{8 \pi x} \, (-2)\mathfrak{Re} \left[ \dfrac{\mu}{M_N} \text{Tr} \left( \mathcal{Y}_\Delta \mathcal{Y}_\nu \mathcal{Y}_\nu^\mathsf{T} \right) \right] \, \dfrac{x - r^2 + 3 a_L}{x - a_L} \Bigg[ \text{ln} |y - a_H| \Bigg]_{y_-}^{y_+}  \ , \\
\widehat{\sigma}_{tt}(x) &= - \dfrac{1}{8 \pi x}  \, \left( \dfrac{\mu}{M_N} \right)^2 \text{Tr} \left( \mathcal{Y}_\nu \mathcal{Y}_\nu^\dagger \right) \, \Bigg[ \dfrac{1 + a_L - a_H}{y - a_H} + \text{ln} | y - a_H| \Bigg]_{y_-}^{y_+} \ . 
\end{align}

\subsection{$\Delta L = 2$ processes} \label{eq:app:deltal2}

There are two $\Delta L = 2$ processes mediated by heavy RHN or the triplet. Each of these processes given in figure \ref{fig:LH_to_LbarHbar} and \ref{fig:LL_to_HbarHbar} consists of three Feynman diagrams in this hybrid scenario whereas the number of diagrams reduces if one considers vanilla seesaw frameworks. It is important to consider only the off-shell part of the heavy unstable propagators in order to avoid double counting \cite{Kolb:1979qa, Giudice:2003jh, Pilaftsis:2003gt, Ala-Mattinen:2023rbm}. We adopt the following conventions for the Breit-Wigner propagator,
\begin{equation}
P^{-1}(x, a, b) = \dfrac{1}{x - a + i \sqrt{b}} \ ; \quad \left| D^{-1}_\text{sub} (x, a, b) \right|^2 = \left| P^{-1}(x, a, b) \right|^2 - \dfrac{\pi}{\sqrt{b}} \delta(x - a) \ ,
\end{equation}
where $\delta(x)$ is the Dirac delta function. The detailed expressions of the reduced cross section of these processes are given below. \\

\noindent
$\bullet$ \textbf{Process:} $L \, (p_1) + H \, (p_2) \rightarrow \overline{L} \, (p_3) + \overline{H} \, (p_4)$ 

The process consists of three Feynman diagrams, RHN mediated s-channel (figure \ref{fig:LH_to_LbarHbar_N_s}) and u-channel (figure \ref{fig:LH_to_LbarHbar_N_u}) diagram and triplet mediated t-channel (\ref{fig:LH_to_LbarHbar_D_t}) diagram for which the transition amplitude is denoted my $\mathcal{M}_s$, $\mathcal{M}_u$ and $\mathcal{M}_t$ respectively. The squared amplitude and the corresponding reduced cross section can be written as 
\begin{align}
|\mathcal{M}|^2(y) &= |\mathcal{M}_s|^2 + |\mathcal{M}_u|^2 + |\mathcal{M}_t|^2 \nonumber \\
&+ \left( \mathcal{M}_s \mathcal{M}_u^\dagger + \mathcal{M}_s^\dagger \mathcal{M}_u \right) + \left( \mathcal{M}_u \mathcal{M}_t^\dagger + \mathcal{M}_u^\dagger \mathcal{M}_t \right) + \left( \mathcal{M}_t \mathcal{M}_s^\dagger + \mathcal{M}_t^\dagger \mathcal{M}_s \right) \ , \\
\widehat{\sigma}(x) &= \int_{y_-}^{y^+} |\mathcal{M}|^2(y) dy = \widehat{\sigma}_{ss}(x) + \widehat{\sigma}_{uu}(x) + \widehat{\sigma}_{tt}(x) + \widehat{\sigma}_{su}(x) + \widehat{\sigma}_{ut}(x) + \widehat{\sigma}_{ts}(x) \ ,
\end{align}
where the explicit expressions of various quantities are given below
%
\begin{align}
x_\text{min} &= (\sqrt{a_L} + \sqrt{a_H})^2 \ , \\
y_- &= - x \lambda(1, a_L/x, a_H/x) \text{ and } y_+ = 0 \ , \\
\widehat{\sigma}_{ss}(x) &=  \dfrac{1}{8 \pi x} \left( \text{Tr} \left( \mathcal{Y}_\nu \mathcal{Y}_\nu^\dagger \right) \right)^2 \left| D^{-1}_\text{sub} (x, 1, a_{\Gamma_N}) \right|^2 \Bigg[ 2 a_L y - \dfrac{y^2}{2} \Bigg]_{y_-}^{y_+} \ , \\
\widehat{\sigma}_{uu}(x) &= \dfrac{1}{8 \pi x} \left( \text{Tr} \left( \mathcal{Y}_\nu \mathcal{Y}_\nu^\dagger \right) \right)^2  \Bigg[ \dfrac{D}{\sqrt{a_{\Gamma_N}}} \tan^{-1} \dfrac{y + C}{\sqrt{a_{\Gamma_N}}} - \dfrac{1}{2} \text{ln} \left| (y + C)^2 + a_{\Gamma_N} \right| \Bigg]_{y_-}^{y_+}      \ , \\
\widehat{\sigma}_{tt}(x) &= -\dfrac{1}{8 \pi x} \left( \dfrac{\mu}{M_N} \right)^2 \text{Tr} \left( \mathcal{Y}_\Delta \mathcal{Y}_\Delta^\dagger \right) \ \times \nonumber \\  
&\hspace{3cm} \Bigg[ \dfrac{1}{2} \text{ln} \left| (y - r^2)^2 + r^2 a_{\Gamma_\Delta} \right| + \dfrac{r^2 - 4a_L}{r \sqrt{a_{\Gamma_\Delta}}} \tan^{-1} \dfrac{y - r^2}{r \sqrt{a_{\Gamma_\Delta}}}  \Bigg]_{y_-}^{y_+}, \\
\widehat{\sigma}_{su}(x) &= \dfrac{\left( \text{Tr} \left( \mathcal{Y}_\nu \mathcal{Y}_\nu^\dagger \right) \right)^2 }{8 \pi x} \dfrac{2 (x -1)}{ \left| P(x, 1, a_{\Gamma_N}) \right|^2 } \ \times \nonumber \\
&\hspace{3cm} \Bigg[y - \sqrt{a_{\Gamma_N}} \tan^{-1} \dfrac{y + C}{\sqrt{a_{\Gamma_N}}} - \dfrac{D}{2} \text{ln} \left| (y + C)^2 + a_{\Gamma_N} \right|  \Bigg]_{y_-}^{y_+} \ , \\
\widehat{\sigma}_{ut}(x) &= \dfrac{1}{8 \pi x} \, 2 \mathfrak{Re} \left\lbrace \dfrac{\mu}{M_N} \text{Tr} \left( \mathcal{Y}_\nu \mathcal{Y}_\nu^\mathsf{T} \mathcal{Y}_\Delta \right)^* \Bigg[ \dfrac{E + F}{F - G} \text{ln} |y + F| - \dfrac{E + G}{F - G} \text{ln} |y + G| \Bigg]_{y_-}^{y_+} \right\rbrace      \ , \\
\widehat{\sigma}_{ts}(x) &= - \dfrac{1}{8 \pi x} \, 2 \mathfrak{Re} \left\lbrace \text{Tr} \left( \mathcal{Y}_\Delta \mathcal{Y}_\nu \mathcal{Y}_\nu^\mathsf{T} \right) P^{-1}(x, 1, a_{\Gamma_N}) \Bigg[ y + (-G^* - E) \text{ln} |y + G^*| \Bigg]_{y_-}^{y_+} \right\rbrace       \ , 
\end{align}
%
with 
\begin{equation} \label{eq:app:const_list}
\begin{aligned}
C &= x + 1 - 2a_L - 2a_H \ , \\
D &= x + 1 - 2 a_H \ , \\
E &= 4 a_L \ , \\
F &= x + 1 - 2a_L - 2a_H - i \sqrt{a_{\Gamma_N}} \ , \\
G &= - r^2 - i r \sqrt{a_{\Gamma_\Delta}} \ .
\end{aligned}
\end{equation}

\noindent
$\bullet$ \textbf{Process:} $L \, (p_1) + L \, (p_2) \rightarrow \overline{H} \, (p_3) + \overline{H} \, (p_4)$ 

The process consists of three Feynman diagrams, triplet mediated s-channel (figure \ref{fig:LL_to_HbarHbar_D_s}) and RHN mediated t-channel and u-channel (figure \ref{fig:LL_to_HbarHbar_N_t} and \ref{fig:LL_to_HbarHbar_N_u}) diagrams for which the transition amplitude is denoted my $\mathcal{M}_s$, $\mathcal{M}_t$ and $\mathcal{M}_u$ respectively. The squared amplitude and the corresponding reduced cross section can be written as 
\begin{align}
|\mathcal{M}|^2(y) &= |\mathcal{M}_s|^2 + |\mathcal{M}_t|^2 + |\mathcal{M}_u|^2 \nonumber \\
&+ \left( \mathcal{M}_s \mathcal{M}_t^\dagger + \mathcal{M}_s^\dagger \mathcal{M}_t \right) + \left( \mathcal{M}_t \mathcal{M}_u^\dagger + \mathcal{M}_t^\dagger \mathcal{M}_u \right) + \left( \mathcal{M}_u \mathcal{M}_s^\dagger + \mathcal{M}_u^\dagger \mathcal{M}_s \right) \ , \\
\widehat{\sigma}(x) &= \int_{y_-}^{y^+} |\mathcal{M}|^2(y) dy = \widehat{\sigma}_{ss}(x) + \widehat{\sigma}_{tt}(x) + \widehat{\sigma}_{uu}(x) + \widehat{\sigma}_{st}(x) + \widehat{\sigma}_{tu}(x) + \widehat{\sigma}_{us}(x) \ ,
\end{align}
where the explicit expressions of various quantities are given below
%
\begin{align}
x_\text{min} &= 4 a_H \ , \\
y_\pm &= - \dfrac{x}{4} \left( \sqrt{1 - \dfrac{4a_L}{x}} \mp \sqrt{1 - \dfrac{4a_H}{x}} \right)^2 \ , \\
\widehat{\sigma}_{ss}(x) &= \dfrac{1}{8 \pi} \left( \dfrac{\mu}{M_N} \right)^2 \text{Tr} \left( \mathcal{Y}_\Delta \mathcal{Y}_\Delta^\dagger \right) \sqrt{1 - \dfrac{4a_L}{x}} \sqrt{1 - \dfrac{4a_H}{x}} 2(x - 4 a_L) \left| D^{-1}_\text{sub} (x, r^2, r^2 a_{\Gamma_\Delta}) \right|^2  \ , \\
\widehat{\sigma}_{tt}(x) &= \dfrac{1}{8 \pi x} \left( \text{Tr} \left( \mathcal{Y}_\nu \mathcal{Y}_\nu^\dagger \right) \right)^2 \dfrac{x - 2a_L}{\sqrt{a_{\Gamma_N}}} \Bigg[ \tan^{-1} \dfrac{y - 1}{\sqrt{a_{\Gamma_N}}} \Bigg]_{y_-}^{y_+}       \ , \\
\widehat{\sigma}_{uu}(x) &= \dfrac{1}{8 \pi x} \left( \text{Tr} \left( \mathcal{Y}_\nu \mathcal{Y}_\nu^\dagger \right) \right)^2 \dfrac{x - 2a_L}{\sqrt{a_{\Gamma_N}}} \Bigg[ \tan^{-1} \dfrac{y + C}{\sqrt{a_{\Gamma_N}}} \Bigg]_{y_-}^{y_+}      \ , \\
\widehat{\sigma}_{st}(x) &= \dfrac{1}{8 \pi x} 2 \mathfrak{Re} \left\lbrace \dfrac{\mu}{M_N} \text{Tr} \left( \mathcal{Y}_\Delta \mathcal{Y}_\nu^* \mathcal{Y}_\nu^\dagger \right)^* \dfrac{x - E}{x - H} \Bigg[ \text{ln} | y + H | \Bigg]_{y_-}^{y_+} \right\rbrace     \ , \\
\widehat{\sigma}_{tu}(x) &= \dfrac{1}{8 \pi x} 2 \mathfrak{Re} \left\lbrace \left( \text{Tr} \left( \mathcal{Y}_\nu \mathcal{Y}_\nu^\dagger \right) \right)^2 \dfrac{x - 2 a_L}{x + 2(1 - a_L - a_H)} \Bigg[ -\text{ln} | y + H^* | + \text{ln} | y + F^*|  \Bigg]_{y_-}^{y_+} \right\rbrace       \ , \\
\widehat{\sigma}_{us}(x) &= \dfrac{1}{8 \pi x} \, 2 \mathfrak{Re} \left\lbrace \dfrac{\mu}{M_N} \text{Tr} \left( \mathcal{Y}_\nu \mathcal{Y}_\nu^\mathsf{T} \mathcal{Y}_\Delta^\dagger \right) \dfrac{x - E}{x + G} \Bigg[ \text{ln} | y + G|  \Bigg]_{y_-}^{y_+} \right\rbrace  \ ,
\end{align}
%
with $H = -1 - i \sqrt{a_{\Gamma_N}}$ and other expressions same as given in eq. \ref{eq:app:const_list}.

\end{appendix}

\bibliographystyle{JHEP}
\bibliography{hybrid_lepto}

\end{document}